\documentclass[aps,prl,superscriptaddress,showpacs,floatfix,nofootinbib,notitlepage,twocolumn,longbibliography]{revtex4-2}
\usepackage{amsmath,graphicx,float,csquotes,mdframed,appendix,url}
\usepackage[dvipsnames]{xcolor}
\usepackage[colorlinks=true, pdfstartview=FitV, linkcolor=RedOrange, citecolor=Emerald, urlcolor=Cerulean]{hyperref}
\usepackage{amssymb}
\usepackage{soul}
\usepackage[normalem]{ulem}

\newcommand{\snn}{\sqrt{s_\mathrm{NN}}}
\newcommand{\plead}{$p$+$^{208}$Pb}
\newcommand{\glb}{\textsc{3d-glauber}}

\begin{document}

\preprint{}

\title{Neutron Skin from Conserved Charge Measurements at Collider Experiments}

\author{Gr\'egoire Pihan}
\email{gpihan@uh.edu}
\affiliation{Physics Department, University of Houston, Box 351550, Houston, Texas 77204, USA}

\author{Akihiko Monnai}
\affiliation{Department of General Education, Faculty of Engineering, Osaka Institute of Technology, Osaka 535-8585, Japan}

\author{Bj\"orn Schenke}
\affiliation{Physics Department, Brookhaven National Laboratory, Upton, New York 11973, USA}

\author{Chun Shen}
\affiliation{Department of Physics and Astronomy, Wayne State University, Detroit, Michigan 48201, USA}

\begin{abstract}
We propose a novel method for measuring the neutron skin of heavy nuclei using collider experiments. Specifically, we demonstrate that the neutron skin thickness of the lead nucleus can be extracted in $p$+$^{208}$Pb collisions by analyzing a double ratio: The ratio of net electric charge to net baryon number measured in the lead-going direction, taken for high-multiplicity events and divided by the same ratio for low-multiplicity events.
We compute the expected sensitivity of the double ratio to the neutron skin within a comprehensive (3+1)D relativistic hydrodynamic framework that incorporates multiple conserved charge currents and a charge-dependent lattice-QCD-based equation of state. We provide predictions for both $p$+$^{208}$Pb collisions at ${\snn=72}$~GeV and $\snn=5.02$~TeV, corresponding to the center of mass energies realized in the SMOG2 fixed-target setup at LHCb and the LHC collider mode, respectively. 
\end{abstract}

\maketitle

Precise measurements of the neutron skin thickness of heavy nuclei, defined as the difference between the neutron and proton root-mean-square (RMS) radii are highly desirable, as they provide constraints on properties of nuclear matter~\cite{Chen:2010qx, Lovato:2022vgq, Sorensen:2023zkk}. This includes the density dependence of the nuclear symmetry energy---an essential piece of the equation of state (EOS) that fixes the mass--radius relation of neutron stars~\cite{Drischler:2020hwi}.

While the charge RMS radius, and by extension, the RMS radius of the proton distribution inside the nucleus ($R_p$), can be determined via elastic electron scattering experiments~\cite{Fricke:1995zz}, extracting the RMS radius of the neutrons ($R_n$), and thereby the neutron--skin thickness, $\Delta R_\mathrm{np} \equiv R_n - R_p$, remains significantly more challenging. The following methods have been used to extract the neutron skin (see \cite{Thiel:2019tkm} for an overview): Parity-violating electron scattering, as realized in the PREX \cite{PREX:2021umo} and CREX \cite{CREX:2022kgg} programs, hadronic scattering at intermediate energies with protons \cite{Zenihiro:2010zz}, alpha particles or pions \cite{Gils:1980zza,Friedman:2012pa}, anti-protonic atoms \cite{Trzcinska:2001sy}, electric–dipole polarizability and the pygmy dipole resonance strength \cite{Tamii:2011pv,Poltoratska:2012nf}, obtained from high‐resolution inelastic proton or photon scattering, and measuring angular distributions in coherent $\pi^{0}$ photoproduction \cite{Tarbert:2013jze}. 

It is noteworthy that the parity-violating electron scattering technique, which provides a model independent determination of the neutron radius of ${}^{208}\rm Pb$, yields a comparatively large neutron skin thickness of $\Delta R_{\rm np}[^{208}\mathrm{Pb}] =0.283\pm0.071\,{\rm fm}$, as reported by PREX-II~\cite{PREX:2021umo}. This value exceeds those obtained from other models by about 1.5$\sigma$ \cite{Hu:2021trw}. A large neutron skin thickness, such as that reported by PREX-II, points to a stiff symmetry energy near saturation density and may, if this stiffness persists at higher densities, favor an earlier activation of the direct Urca process \cite{Lattimer:1991ib, Thapa:2022zkr} in the cores of massive neutron stars. Neutron skin measurements \cite{NANOGrav:2019jur, Fonseca:2021wxt, Miller:2021qha} therefore play an important role in informing the density dependence of the symmetry energy and its potential impact on neutron-star cooling within unified equation-of-state frameworks.

Recently, it has been demonstrated that the neutron skin can also be constrained through a Bayesian analysis of relativistic nuclear collision data \cite{Giacalone:2023cet}. In such collisions, the spatial distributions of protons and neutrons within the nuclei influence the initial geometry and the number of participant nucleons. In turn they have a sizable effect on final-stage observables such as the total multiplicity, mean transverse momentum, and anisotropic flow coefficients. This method is rather indirect as no information on the electric charge is used. Other recent studies proposed to probe the neutron skin size with free spectator nucleons from heavy-ion collisions~\cite{Liu:2023qeq}, or using the jet charge in deep inelastic scattering off nuclei \cite{Zhang:2025raf}.

In this article, we propose a novel method to constrain the neutron skin of heavy nuclei, such as the $^{208}{\rm Pb}$ nucleus, using measurements of conserved charges at the Large Hadron Collider (LHC) at CERN. Specifically, we construct the double ratio of the rapidity distributions of net electric charge ($Q$) and net baryon number ($B$) (or equivalently with experimental $B$ and $Q$ proxies) for central and peripheral proton-nucleus collisions. The evolution of these charges from the initial collision to freeze-out is governed by the non-linear dynamics of the quantum chromodynamic (QCD) medium \cite{Andersson:1983ia, Vance:1997th, Sjostrand:2006za, Bahr:2008pv, Pratt:2023pee}, making it highly non-trivial. However, we find by explicit calculation that the relevant effects of the neutron skin on computed experimentally observable net-charge rapidity distributions closely resemble those on the corresponding initial state distributions. We demonstrate this in Fig.~\ref{fig:SMsmallr}  of the Supplemental Material. 

In the presence of a neutron skin, low event activity $p$+A collisions involve, on average, more proton-neutron interactions than high event activity collisions, because low event activity is correlated with a large impact parameter, i.e., the proton hitting the edge of the lead nucleus.\footnote{The event activity is defined using the charged particle multiplicity in a given rapidity range. While weaker than in A+A collisions, the correlation of multiplicity with the impact parameter is sizable enough to distinguish between small and large impact parameters. Due to this correlation, we denote event activity classes as centrality classes below.} This alters the relative number of proton and neutron participants and thus affects the distribution of net-electric charge densities in the final stage. We demonstrate that the double ratio exhibits an approximately linear dependence on the neutron skin thickness, establishing it as a sensitive and robust probe of the lead neutron skin in \plead{} collisions. We strongly advocate for the experimental evaluation of this observable in \plead{} collisions, as it offers a novel and rather direct means of measuring the neutron skin of the colliding nuclei.

We evaluate the double ratio observable using state-of-the-art (3+1)D event-by-event hydrodynamic simulations of \plead{} collisions at $\snn = 5.02$~TeV, corresponding to the LHC collider mode - and at $\snn = 72$~GeV, relevant to the System for Measuring Overlap With Gas 2 (SMOG2) fixed-target program at LHCb ($^{208}$Pb on hydrogen). These simulations employ the i\textsc{EBE-MUSIC} framework~\cite{Pihan:2024lxw, pihan_2025_17193964}.  We construct the initial energy density and conserved charge distributions using the \textsc{3D-Glauber} model, which builds on the Monte Carlo Glauber approach to construct the geometry in the transverse plane and computes a longitudinal structure through decelerating strings formed between wounded nucleons~\cite{Shen:2017bsr, Shen:2022oyg}. 

To describe the nuclear density profiles and account for the neutron skin, we apply an isospin-dependent Woods–Saxon parametrization to sample the nucleon positions inside the colliding nuclei. The proton and neutron density distributions are defined as 
\begin{equation}
    \rho_{p,n}(r) = 
    \frac{1}{1+e^{(r - R_{p,n}^{\rm WS})/a_{p,n}}}\,.
    \label{eq:WSprofile1}
\end{equation}
Assuming spherical symmetry for the lead nucleus, the proton and neutron density distributions, $\rho_p(r)$ and $\rho_n(r)$, are isotropic and depend only on the radial coordinate $r$. The half-density radii of the proton and neutron distributions are set to $R_p^{\rm WS} = 6.68$~fm and $R_n^{\rm WS} = 6.69$~fm as in Ref~\cite{Giacalone:2023cet,Zhang:2025raf}, in line with the halo-type neutron skin. 

While the diffuseness of the proton distribution is kept constant at $a_p = 0.448$\,fm \cite{Trzcinska:2001sy,Zenihiro:2010zz}, we vary the neutron diffuseness by introducing a relative parameter $\Delta a_\mathrm{np} = a_n - a_p$, allowing for a controlled adjustment of the neutron skin thickness in the lead nucleus. The value of $\Delta a_\mathrm{np}$ affects the shape of the Woods-Saxon distribution \eqref{eq:WSprofile1} and consequently the neutron skin thickness $\Delta R_\mathrm{np}$. We will use it as a measure of the neutron skin in the remainder. For the specific case of a vanishing neutron skin thickness $\Delta R_{\text{np}} = 0.0$~fm, we use $R^{\text{WS}}_n = R^{\text{WS}}_p = 6.68$~fm and $a_n = a_p = 0.448$~fm.  

The \glb~model allows for differences in baryon charge and electric charge stopping, their transport toward midrapidity in a collision. One motivation for such a difference is the existence of baryon junctions \cite{Montanet:1980te} whose dynamics were studied in Refs.~\cite{Kharzeev:1996sq, Shen:2022oyg, Pihan:2024lxw}. Here, baryon charges are assigned to string junctions that are stopped more easily than valence quarks (as junctions are identified with gluons, their initial longitudinal momentum fractions are lower), shifting the baryon charges closer to midrapidity. For simplicity, we assume that the parameter that controls the difference between baryon and electric charge stopping remains the same across all collision energies, although a more rigorous treatment would account for its energy dependence, which can be constrained by experimental data. The working value for this parameter has been constrained from Au+Au data at $\snn=200$~GeV \cite{Pihan:2024lxw} hence, suited for fixed target mode. For the collider mode, considering the expected dependence as discussed in Ref~\cite{Kharzeev:1996sq}, the parameter would be smaller than our working value. As we will include the case for equal baryon and electric charge stopping in the evaluation of the systematic error, this effect will be taken into account in our final results. We emphasize that the existence of baryon junctions is in no way required for the proposed method of determining the neutron skin. In the Supplemental Material, we discuss that its effects are negligible at very forward rapidities, while different stopping mechanisms introduce systematic uncertainties for smaller rapidities, which we include in the presented results.

After constructing the initial distributions of energy density and conserved charges, we initialize the hydrodynamic evolution at a fixed proper time $\tau_\mathrm{hydro} = 0.5$~fm/$c$. 
The system then evolves according to the conservation laws~\cite{Schenke:2010nt, Schenke:2010rr, Paquet:2015lta, Denicol:2018wdp, Pihan:2023dsb, Pihan:2024lxw}: 
\begin{align}
    \partial_{\mu} T^{\mu \nu} &= 0, \text{~~and~~}
    \partial_{\mu} J_{B,Q}^{\mu} = 0,
    \label{Eq:idealHydroEoM}
\end{align}
where $T^{\mu \nu} = \varepsilon u^\mu u^\nu + \mathcal{P} (u^\mu u^\nu -g^{\mu\nu}) +\Pi^{\mu\nu}$ is the energy-momentum tensor and ${J_{B,Q}^{\mu} \equiv n_{B,Q} u^{\mu}}$ denotes the conserved charge currents. In the energy-momentum tensor, $\varepsilon$ is the energy density, $\mathcal{P}$ the pressure, $u^{\mu}$ the fluid 4-velocity, and $\Pi^{\mu \nu}$ the viscous stress tensor. For the conserved charge currents, we do not include a dissipative part.  While diffusion of conserved charges can influence their longitudinal evolution during hydrodynamic expansion, we neglect these effects in this study. As the double ratio observable is constructed through the comparison of different centralities, in which diffusion effects on the longitudinal distributions are similar, we do not observe a large impact from the diffusion of the conserved charges (see an explicit study in the Supplemental Material).

We include bulk and shear viscous effects in the hydrodynamic evolution by solving the Denicol–Niemi–Molnar–Rischke (DNMR) second-order viscous hydrodynamic equations \cite{Denicol:2012cn, Denicol:2018wdp}. The specific shear and bulk viscosities are parametrized as~\cite{Pihan:2023dsb, Pihan:2024lxw}
\begin{align}
    \frac{\eta T}{\varepsilon + \mathcal{P}} &= \eta_0 \left[1 + b \left(\frac{\mu_B}{\mu_{B, 0}}\right)^a \right] \,,\\
    \frac{\zeta T}{\varepsilon + \mathcal{P}} &= \zeta_0 \exp \left[-\left(\frac{T - T_\mathrm{peak}}{T_{\mathrm{width}, \lessgtr}}\right)\right]\,,
\end{align}
where $T_\mathrm{peak} = 0.17\,\mathrm{GeV} - 0.15\;\mathrm{GeV}^{-1} \mu_B^2$ denotes the temperature at which the bulk viscosity peaks. The widths of the Gaussian profile are $T_{\mathrm{width},<} = 0.01$~GeV for $T < T_\mathrm{peak}$ and $T_{\mathrm{width},>} = 0.08$~GeV for $T > T_\mathrm{peak}$. We use $\zeta_0 = 0.1$ for the peak value of bulk viscosity, and set the shear viscosity parameters to $\eta_0 = 0.08$, $b = 2$, $\mu_{B,0} = 0.6$~GeV, and $a = 0.7$~\cite{Shen:2020jwv}. We use the \textsc{neos}-4D equation of state \cite{Monnai:2024pvy, Monnai:2025nyg}, which constructs the pressure as a function of the temperature and chemical potentials of baryon, electric charge, and strangeness.

Once the local energy density in a fluid cell falls below the threshold $\varepsilon_\mathrm{sw} = 0.2$ GeV/fm$^3$, we identify the corresponding constant energy density hypersurface using the Cornelius algorithm~\cite{Huovinen:2012is} and convert the fluid into hadrons via the Cooper–Frye prescription~\cite{Cooper:1974mv, Shen:2014vra}. Sampled hadrons are then propagated through the hadronic phase using the UrQMD transport model~\cite{Bass:1998ca, Bleicher:1999xi}, which accounts for hadronic rescatterings and decays. Both strong and weak decays are included in constructing the final-state observables of stable hadrons.

We define the double ratio observable as
\begin{equation}
    \mathcal{R}^{X, Y}_{c_1, c_2}(y_1, y_2) = \frac{N_X(y_1, y_2, c_1)}{N_Y(y_1, y_2, c_1)} \bigg/ \frac{N_X(y_1, y_2, c_2)}{N_Y(y_1, y_2, c_2)} ,
    \label{eq:Rdef}
\end{equation}
where $X$ and $Y$ represent the types of conserved charges or their experimental proxy. The quantity $N_X(y_1, y_2, c)$ is defined as
\begin{equation}
    N_X(y_1, y_2, c) = \left\langle \int_{y_1}^{y_2} \mathrm{d}y \, \frac{d N_X}{dy}(y, c)  \right\rangle_{\mathrm{ev}},
\end{equation}
with $c$ denoting the centrality class and $\langle \cdot \rangle_{\mathrm{ev}}$ indicating an average over all events in the given centrality. Throughout this work, we focus on the case {$c_1 = 80$-$100\%$} (peripheral collisions) and {$c_2 = 0$-$20\%$} (central collisions). We consider two specific choices for the double ratio defined in Eq.~\eqref{eq:Rdef}:
\begin{equation*}
    \bullet\,\,\mbox{Case (1): } X=Q, Y=B, \quad \bullet\,\,\mbox{Case (2): } X=Q, Y=p
\end{equation*}
where $Q$ is the net electric charge (computed from charged pions, kaons, and protons), $B$ the net baryon number (computed from protons and neutrons) and $p$ the net proton number. In case (1), $\mathcal{R}^{QB}_{c_1, c_2}$ captures the change in the ratio of net electric charge to net baryon number between peripheral and central collisions. Case (2) is an experimentally accessible proxy, where net protons stand in for net baryons, because net neutrons are challenging to measure.

In both scenarios, a finite neutron skin reduces the double ratio. In peripheral collisions, the neutron-rich surface layer depletes proton participants, lowering the net electric charge relative to the net baryon number. Central collisions, in contrast, are less sensitive to the neutron skin and thus provide a natural baseline. Consequently, the double ratio $\mathcal{R}^{X, Y}_{c_1, c_2}$ decreases with increasing neutron skin thickness and approaches unity when the neutron skin becomes negligible. 

We present results for two center-of-mass energies: $\snn=5.02$~TeV, corresponding to LHC \textit{collider mode}, and $\snn = 72$~GeV, corresponding to the LHCb SMOG2 \textit{fixed target} experiment. Centrality classes are defined using the charged-particle multiplicity distributions in the rapidity intervals $1.25 < y^{\rm lab} < 3$ for collider mode and $0.6 < y^{\rm lab} < 2.6$ for the fixed target setup. These intervals are chosen to avoid spurious correlations by ensuring that the centrality selection does not overlap with the rapidity regions used in the analysis. For the analysis, we consider two laboratory-frame rapidity intervals: $4.5 < y^{\rm lab} < 5.5$, corresponding to the \textit{forward} acceptance at LHCb, and $5.5 < y^{\rm lab} < 7.5$, located deeper in the \textit{fragmentation} region, where sensitivity to the neutron skin is expected to be cleanest, though this region may not yet be experimentally accessible. We note that in collider mode, since the rapidity shift is small compared to the beam rapidity, studying the very forward region is pushing the applicability of the hydrodynamics to its limit. However, we focus exclusively on ratios of conserved charges integrated in the full space. As such the only important effect from the hydrodynamic is the global charge conservation. Local dynamical details of the hydrodynamic evolution, while potentially relevant for differential observables, have only a subleading impact here. Hydrodynamics therefore provides a robust framework for our study.

\begin{figure}[!ht]
    \includegraphics[width=0.45\textwidth]{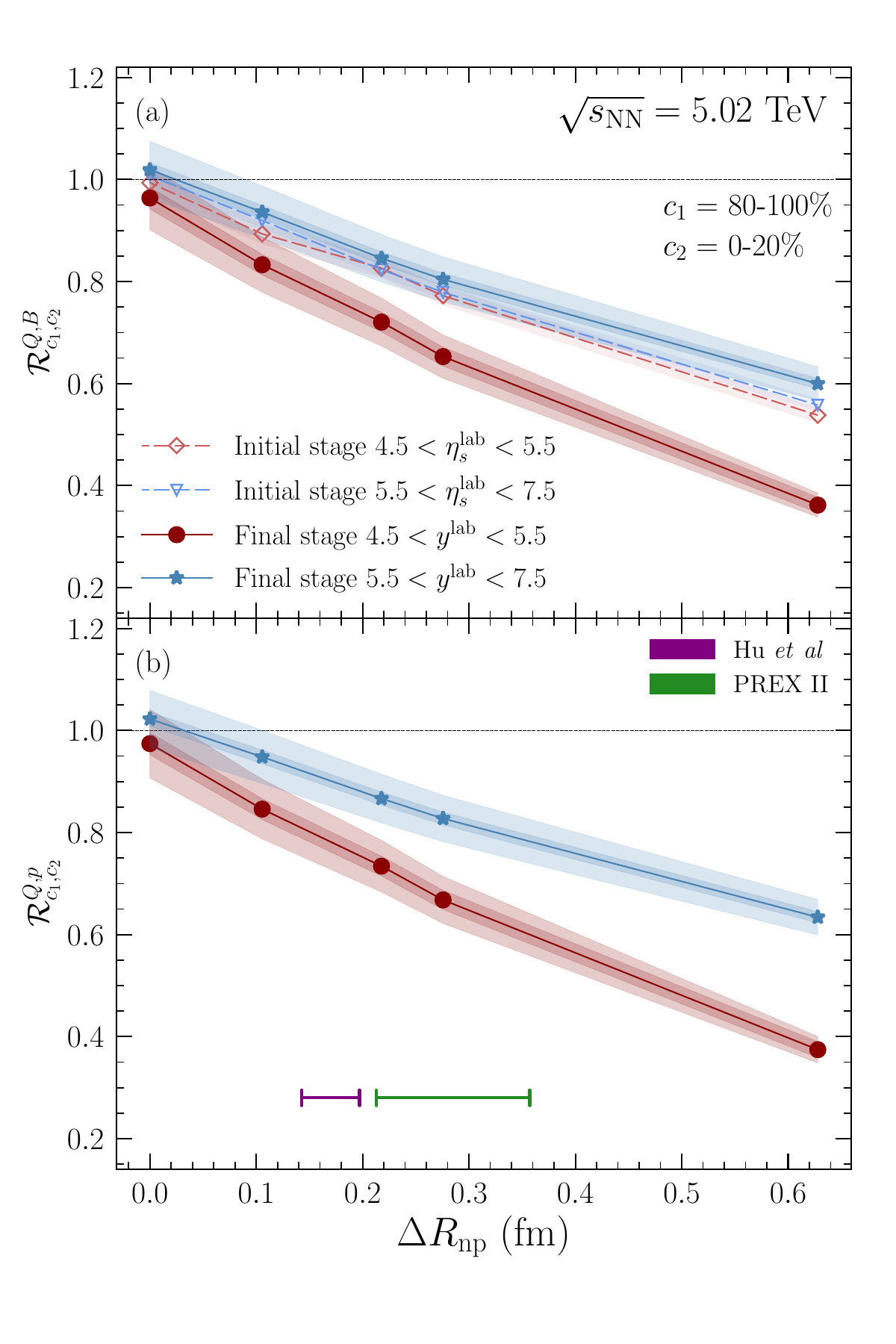}
    \caption{
    The double ratio as a function of the neutron skin thickness RMS in the initial (dashed lines) and final stage (solid lines) in $\snn=5.02$~TeV \plead{} collisions. Panel (a): Case (1) measured in $4.5 < \eta_s^{\rm lab}, y^{\rm lab} < 5.5$ (diamond and circle markers) and in  $5.5 < \eta_s^{\rm lab}, y^{\rm lab} < 7.5$ (triangle and star markers). Panel (b): Case (2) measured in $4.5 < \eta_s^{\rm lab}, y^{\rm lab} < 5.5$ (circle markers) and in  $5.5 < \eta_s^{\rm lab}, y^{\rm lab} < 7.5$ (star markers). For final stage results, the darker error bands represent statistical errors and the light error bands the systematic errors.
    The purple band shows the neutron skin RMS result from Hu \textit{et al.}~\cite{Hu:2021trw} and the green band is the PREX II measurement \cite{PREX:2021umo}. 
    }
    \label{fig:R502}
\end{figure}

We present the double ratio as a function of the neutron skin thickness, $\Delta R_\mathrm{np}$, in collider mode \plead{} at $\snn = 5.02$~TeV for case (1) in Fig.\,\ref{fig:R502}(a) and for case (2) in Fig.\,\ref{fig:R502}(b).
Panel (a) contrasts initial-stage predictions (dashed lines) with final-stage results (solid lines) in the forward ($4.5 < y^{\rm lab}, \eta_s^{\rm lab} < 5.5$) and fragmentation ($5.5 < y^{\rm lab}, \eta_s^{\rm lab} < 7.5$) rapidity windows. As expected, the double ratio starts out near one for $\Delta R_\mathrm{np} \approx 0$, and decreases monotonically with increasing $\Delta R_\mathrm{np}$. 
In the \emph{fragmentation}
rapidity range, the final state double ratio closely follows the initial state result. In the \emph{forward} region the result is more sensitive to final state effects, which will be further discussed below.
Case (2) in panel (b),  which approximates the baryon number by the proton number, shows nearly identical results to case (1), indicating that the net proton number serves as a robust proxy for the total net baryon number in this measurement. 

\begin{figure}[!t]
    \includegraphics[width=0.45\textwidth]{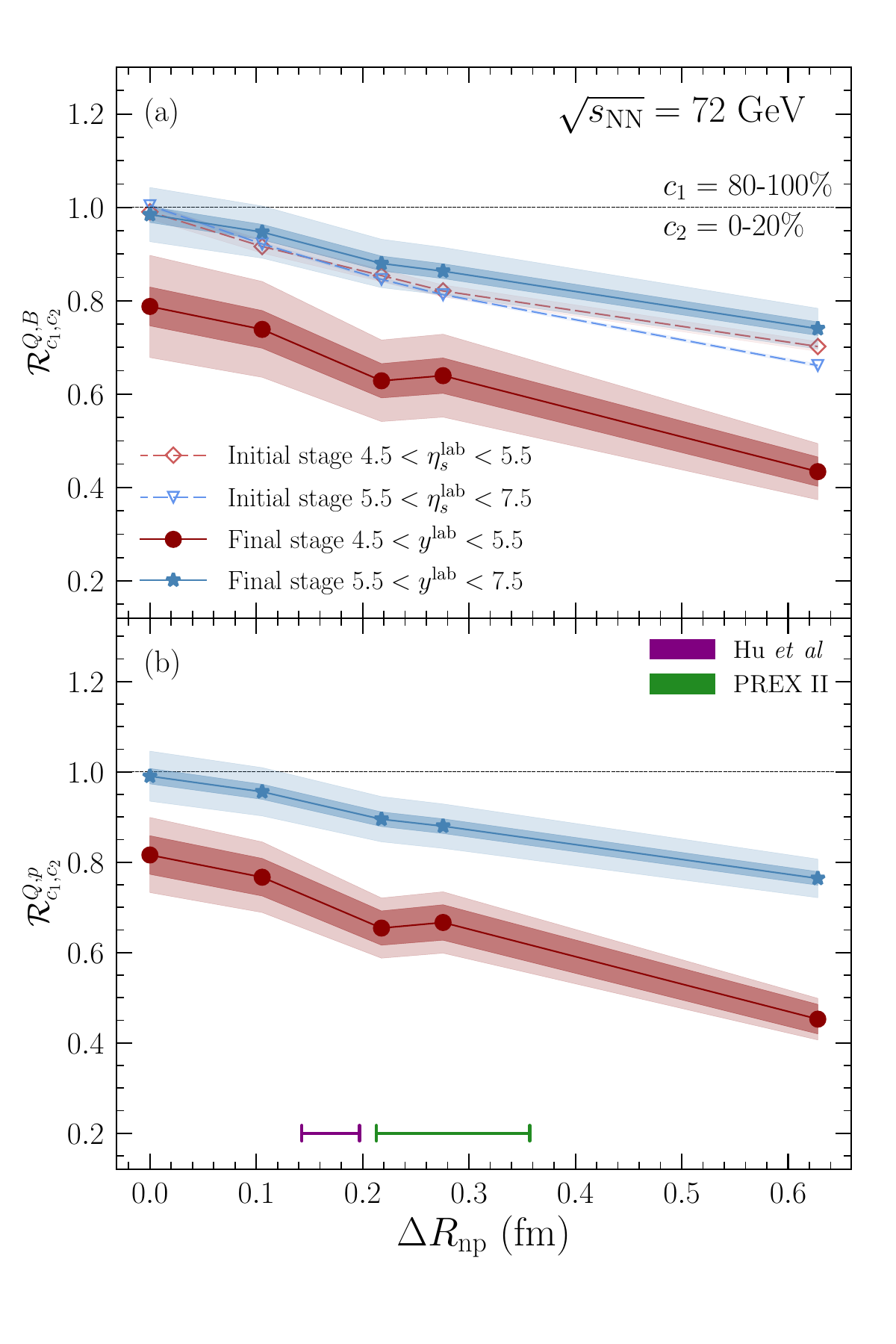}
    \caption{Double ratios as in Fig.~\ref{fig:R502} but for the fixed-target mode \plead{} collisions at $\snn=72$~GeV. For final stage results, the darker error bands represent statistical errors and the light error bands the systematic errors.}
    \label{fig:R72}
\end{figure}

Figure\,\ref{fig:R72} shows that the results for the fixed-target setup are similar to the collider setup. In both figures, the darker error bands represent statistical uncertainties, while the lighter bands indicate systematic ones. The latter arise from three sources. To estimate the uncertainty from unknown details in the baryon stopping, we remove the effect of enhanced baryon stopping (compared to electric charge stopping), which leads to effects below $5$\% ($13$\%) in the forward rapidity bin and below $1.7$\% ($2.6$\%) in the fragmentation region for the collider (fixed-target) mode. Second, the relative diffusion of $B$ and $Q$ during the hydrodynamic evolution is estimated from 1+1D hydrodynamics calculations including diffusion currents for Au+Au at $62.4$~GeV. It yields a maximal $5$\% effect on the final double ratio and we assume a similar $5\%$ effect in \plead{} collisions for both energies. Third, the neutron-skin thickness is modeled through the isospin dependence of the Woods–Saxon diffuseness parameter. Instead, varying the neutron skin by changing the half-density radius at fixed diffuseness introduces small differences in the double ratio. The relative systematic errors on the double ratios from the method of modeling the neutron skin have been estimated for realistic experimental values of the neutron skin thickness $\Delta R_{np}\sim 0.1-0.3$~fm. We use the values quoted in tables \ref{table:SystErr502} and \ref{table:SystErr72} in the Supplemental Material for all $\Delta R_{np}$.

For fixed target collisions, we find that in the experimentally accessible \textit{forward} rapidity bin, the double ratio deviates from unity by approximately $20\%$ for vanishing neutron skin (for collider mode the result is consistent with unity within errors). 

This deviation comes from the species-dependent thermal/momentum smearing at particlization: the switching surface preserves net charge–baryon balance, but thermal emission distributes them differently across particle species. Lighter pions get a broader rapidity spread than heavier baryons, so electric charge (pion-dominated) spreads more than baryon number, producing a centrality-dependent shift in $Q/B$ and thus the double ratio. We verified that the double ratio on the particlization surface is consistent with one for vanishing neutron skin and equal stopping for $B$ and $Q$. 

These effects hinder neutron-skin extraction at \emph{forward} rapidity. By contrast, in the \emph{fragmentation} region the double ratio is robust to variations in stopping and particlization, and at both energies approaches 1 as $\Delta R_\mathrm{np}\to 0$. In collider mode, forward rapidity is farther from the center of mass frame midrapidity than in fixed-target mode, reducing sensitivity to these effects.

For reference, we indicate in both Figs.\,\ref{fig:R502} and \ref{fig:R72} the ranges of $\Delta R_\mathrm{np}$ from the calculations of Hu \textit{et al.}\,\cite{Hu:2021trw} (purple) and from the PREX-II measurement \cite{PREX:2021umo} (green). Especially in collider mode, the proposed double ratio provides a clean and sensitive probe of the neutron skin at LHC energies. The observable reacts systematically to variations in $\Delta R_\mathrm{np}$, and can be measured experimentally in the forward region for case (2). Depending on the achievable experimental uncertainty, the measurement in collider mode can have a similar precision for $\Delta R_\mathrm{np}$ as the existing extractions shown.

\begin{figure}[!ht]
    \includegraphics[width=0.45\textwidth]{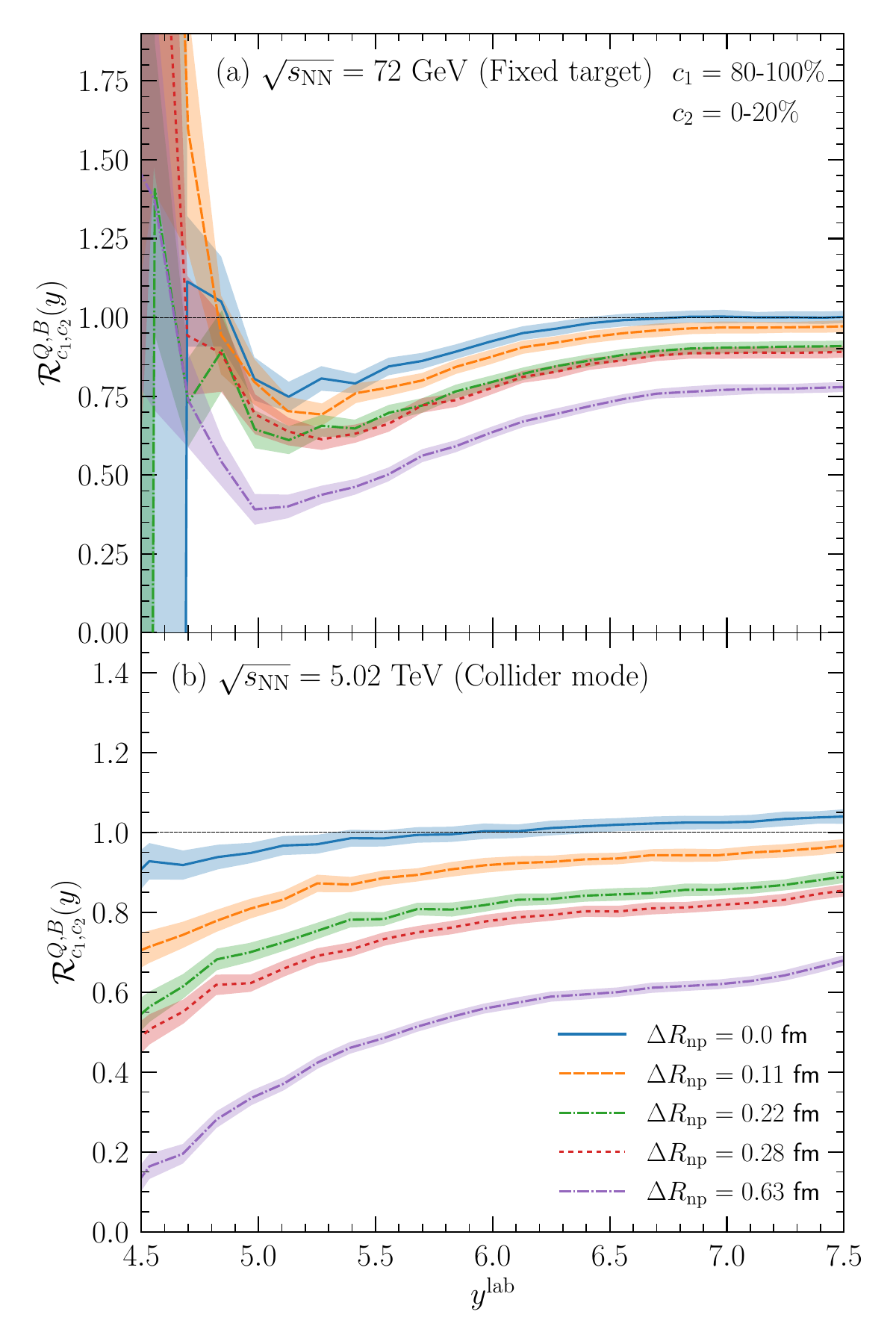}
    \caption{The differential $Q/B$ double ratio in the laboratory frame as a function of the rapidity for different values of the neutron skin thickness at $\snn=72$~GeV (a) and $\snn = 5.02$~TeV (b).
    }
    \label{fig:RapDep}
\end{figure}
To gain additional insight and to highlight that the observable is more robust at larger rapidities, we explore the rapidity dependence of the double ratio $\mathcal{R}^{Q, B}_{c_1, c_2}(y)$ in Fig.\,\ref{fig:RapDep} for various values of the neutron skin thickness and for both collision energies. The rapidity-dependent form of the double ratio is defined as:
\begin{equation}
\!\!\! \mathcal{R}^{Q, B}_{c_1, c_2}(y) \equiv \frac{\langle dN_{Q}/dy\rangle_{\rm ev}(y, c_1)}{\langle dN_{B}/dy\rangle_{\rm ev}(y, c_1)}
\bigg/
\frac{\langle dN_Q/dy\rangle_{\rm ev}(y, c_2)}{\langle dN_B/dy\rangle_{\rm ev}(y, c_2)},
\end{equation}
where $\langle dN_{Q, B}/dy\rangle_{\rm ev}(y, c)$ denotes the event-averaged rapidity distribution of net electric charge and baryons in centrality class $c$, respectively. This expression corresponds to Eq.\,\eqref{eq:Rdef}, evaluated locally in rapidity without prior integration over the rapidity interval.

We find that the overall magnitude of the double ratio strongly depends on the neutron skin thickness $\Delta R_{\rm np}$, with a clear separation between curves at different values. For both collision energies, a nontrivial rapidity dependence can be observed at lower rapidities, while for large rapidities most curves approach a constant. This reaffirms that measurements should be done around as large a rapidity as possible.

We introduced a novel observable to probe the neutron skin thickness of heavy nuclei in high-energy proton–nucleus collisions, based on direct observation of conserved electric and baryon charges. Specifically, we demonstrated that the centrality based double ratio $\mathcal{R}^{X, Y}_{c_1, c_2}$—constructed from net electric charge and net baryon yields—offers direct sensitivity to neutron skin effects in \plead{} collisions at the LHC. 

Because this ratio encodes the relative contributions of neutron and proton participants to particle production, it provides a clear handle on deviations from a uniform nucleon distribution within the nucleus. 

We presented results for two implementations of the double ratio, $Q/B$ and $Q/p$, evaluated in two different forward rapidity intervals. Among these, the net charge-to-proton ratio in the window $4.5 < y < 5.5$ is the most promising candidate for near-term measurements at LHC. While systematic uncertainties associated with particlization and charge stopping may limit direct model-to-data comparisons in the fixed-target configuration at present, the collider-mode predictions are ready for experimental tests.

Such measurements would open a new avenue for constraining the neutron skin of ${}^{208}\mathrm{Pb}$. The same methodology can be applied to other nuclei in high-energy $p+A$ collisions, such as $^{48}$Ca.

The data that support the findings of this article are openly available~\cite{pihan_2025_17193964}, embargo periods may apply.

\paragraph{Acknowledgments}

This work is supported by the U.S. Department of Energy, Office of Science, Office of Nuclear Physics, under DOE Contract No.~DE-SC0012704 and within the framework of the Saturated Glue (SURGE) Topical Theory Collaboration (B.P.S.) and Award No.~DE-SC0021969 (C.S. \& G.P.). This work is also supported by JSPS KAKENHI Grant Number JP24K07030 (A.M.).
C.S. acknowledges a DOE Office of Science Early Career Award. 
This research was done using computational resources provided by the Open Science Grid (OSG)~\cite{Pordes:2007zzb,Sfiligoi:2009cct,OSPool,OSDF}, which is supported by the National Science Foundation awards \#2030508 and \#2323298.

\appendix

\section{Appendix A: Estimation of systematic errors}
\label{sec:AppA}

\begin{table}[ht!]
\centering
\caption{The relative error (\%) obtained for the collider mode ($\snn=5.02$~TeV) for typical values of the neutron skin thickness ($\Delta R_{np} \sim 0.1-0.3$~fm) coming from the diffusion, baryon stopping uncertainty, and prescription for modeling the neutron skin (use of the half density radius of the neutron $R_n^{\rm{WS}}$ as a measure of the neutron skin instead of the diffuseness of the potential). The total relative errors are computed from the quadratic sum of the absolute errors calculated at $\Delta R_{np} \sim 0.22$~fm. 
}
\begin{tabular}{|c|c|c|c||c|}
\hline
 $4.5 < y^{\rm lab} < 5.5$ & Diffusion & Baryon stopping & $R_n^{\rm{WS}}$ & total\\
\hline
$\mathcal{R}^{Q, B}_{c_1, c_2}$ & 5.0 & 4.0 & 1.5 & 6.5 \\
$\mathcal{R}^{Q, p}_{c_1, c_2}$ & 5.0 & 4.5 & 1.5 & 6.9 \\
\hline
\hline
$5.5 < y^{\rm lab} < 7.5$ & Diffusion & Baryon stopping & $R_n^{\rm{WS}}$ & total \\
\hline
$\mathcal{R}^{Q, B}_{c_1, c_2}$ & $5.0$ &  1.7 & 1.6 & 5.5 \\
$\mathcal{R}^{Q, p}_{c_1, c_2}$ & $5.0$   & 1.7 & 1.3 & 5.5\\
\hline
\end{tabular}
\label{table:SystErr502}
\end{table}

\begin{table}[ht!]
\centering
\caption{The relative error (\%) obtained for the fixed target mode ($\snn=72$~GeV) for typical values of the neutron skin thickness ($\Delta R_{np} \sim 0.1-0.3$~fm) coming from the diffusion, baryon stopping uncertainty, and prescription for modeling the neutron skin. The total relative errors are computed from the quadratic sum of the absolute errors calculated at $\Delta R_{np} \sim 0.22$~fm. 
}
\begin{tabular}{|c|c|c|c||c|}
\hline
 $4.5 < y^{\rm lab} < 5.5$ & Diffusion & Baryon stopping & $R_n^{\rm{WS}}$ & total\\
\hline
$\mathcal{R}^{Q, B}_{c_1, c_2}$ & 5.0 & 12.9 & 1.5 & 13.9 \\
$\mathcal{R}^{Q, p}_{c_1, c_2}$ & 5.0 & 8.8 & 1.5 & 10.2 \\
\hline
\hline
$5.5 < y^{\rm lab} < 7.5$ & Diffusion & Baryon stopping & $R_n^{\rm{WS}}$ & total \\
\hline
$\mathcal{R}^{Q, B}_{c_1, c_2}$ & $5.0$ &  2.6 & 1.6 & 5.9 \\
$\mathcal{R}^{Q, p}_{c_1, c_2}$ & $5.0$   & 2.6 & 1.3 & 5.6\\
\hline
\end{tabular}
\label{table:SystErr72}
\end{table}

In Fig.~\ref{fig:R502} and Fig.~\ref{fig:R72} we present systematic uncertainty bands. They are computed by considering the following three major sources of systematic errors: 

\begin{itemize}
    \item \underline{Diffusion effects}: 
    
    During the hydrodynamic evolution, our main result only considers the ideal part of the conserved charge 4-current, $J^{\mu}_{B,Q} = n_{B,Q} u^{\mu}$. 
    Based on the work in Ref.~\cite{Monnai:2026fkp}, where the chosen conductivities are in line with values in Refs.~\cite{Greif:2017byw,Rougemont:2017tlu}, we evaluated the impact of the presence of diffusion of the net baryon and net electric densities in Au+Au collisions at $\snn = 62.4$~GeV. Evaluating the exact same double ratios as in Eq.~\eqref{eq:Rdef} we observed that there is indeed a residual impact of the diffusion. 
    To mimic the effect of the neutron skin, we vary the initial $Q/B$ ratio in the peripheral events, while keeping $Q/B=0.4$ fixed in the central events. We observe that the total deviation of the double ratio caused by diffusion is $\sim1\%$ for initial $Q/B = 0.36$, $\sim3\%$ for $Q/B = 0.32$ and $\sim5\%$ for $Q/B = 0.28$. For typical values of the neutron skin thickness (e.g.~$\Delta R_{\rm np} \sim 0.1-0.3$~fm) the initial $Q/B$ is approximately 0.32 for \plead~collisions, corresponding to an uncertainty of $3\%$. However, to be conservative and account for the fact that transverse gradients were not considered in this estimate, we chose to report an error of 5\%. This is the value reported in tables \ref{table:SystErr502} and \ref{table:SystErr72}.
    
    \item \underline{Modeling of the neutron skin thickness}

    We use a halo-type Wood-Saxon parametrization of the neutron skin thickness $\Delta R_{\rm np}$, where only the diffuseness parameter is isospin dependent. Alternatively, one could also make the half density radius parameter isospin dependent. Evaluating the double ratios for typical values of the neutron skin thickness $\Delta R_{\rm np}\sim0.1-0.3$~fm using the half density radius $R^{\rm WS}_n$ at fixed diffuseness $a_n = a_p=0.01$~fm, we find that the resulting uncertainty on the double ratio is less than $3\%$. We use and report the exact values in tables \ref{table:SystErr502} and \ref{table:SystErr72}.

    \item \underline{Baryon stopping}

    Our study incorporates a model of increased baryon over electric charge stopping, constrained by experimental data.
    One interpretation of why baryons could be stopped more strongly is the existence of a baryon junction. The parameter $P(B_s)$ quantifies the probability for the baryon number of a wounded nucleon to be transported towards midrapidity, according to the distribution of baryon junctions after the collision derived in \cite{Kharzeev:1996sq}. If one choose $P(B_s) = 0$, the baryon stopping comes solely from the valence quark stopping, if $P(B_s)$ is finite, the valence quark stopping is supplemented by a junction contribution.
    
    In the present study we use $P(B_s) = 0.1$, which leads to enhanced baryon stopping as in Ref.~\cite{Pihan:2024lxw}, and is thus favored by experimental data at RHIC energies. To evaluate the systematic impact of uncertainties in the stopping mechanism, we also estimate the double ratio for the extreme case of $P(B_s) = 0$. For collider mode, shown in  table \ref{table:SystErr502}, we see that the relative error is less than 4.5\% for the $4.5 < y^{\rm lab} < 5.5$ rapidity range and less than 2\% for $5.5 < y^{\rm lab} < 7.5$. For the fixed target mode, Table \ref{table:SystErr72}, the error is larger, especially in the forward acceptance rapidity range $4.5 < y^{\rm lab} < 5.5$, where the systematic error is up to 13\%. 
\end{itemize}

The total systematic uncertainties in the column "total" in tables \ref{table:SystErr502} and \ref{table:SystErr72} are compute from the quadratic sum (independent error) of the absolute errors evaluated at $\Delta R_{\rm np} = 0.22$~fm, corresponding to the neutron skin thickness quoted in Ref.~\cite{Hu:2021trw}. These relative uncertainties are shown by the systematic error bands in Figs.~\ref{fig:R502} and \ref{fig:R72}. 

\section{Survival of the neutron skin effect from initial stage to final stage}
\label{B}

The depletion of the net electric charge due to a finite neutron skin thickness in peripheral compared to central collision is an initial state effect. The neutron skin affects the B and Q deposition in the medium at the earliest times of the collision. 
\begin{figure}[!ht]
    \centering
    \includegraphics[width=0.45\textwidth]{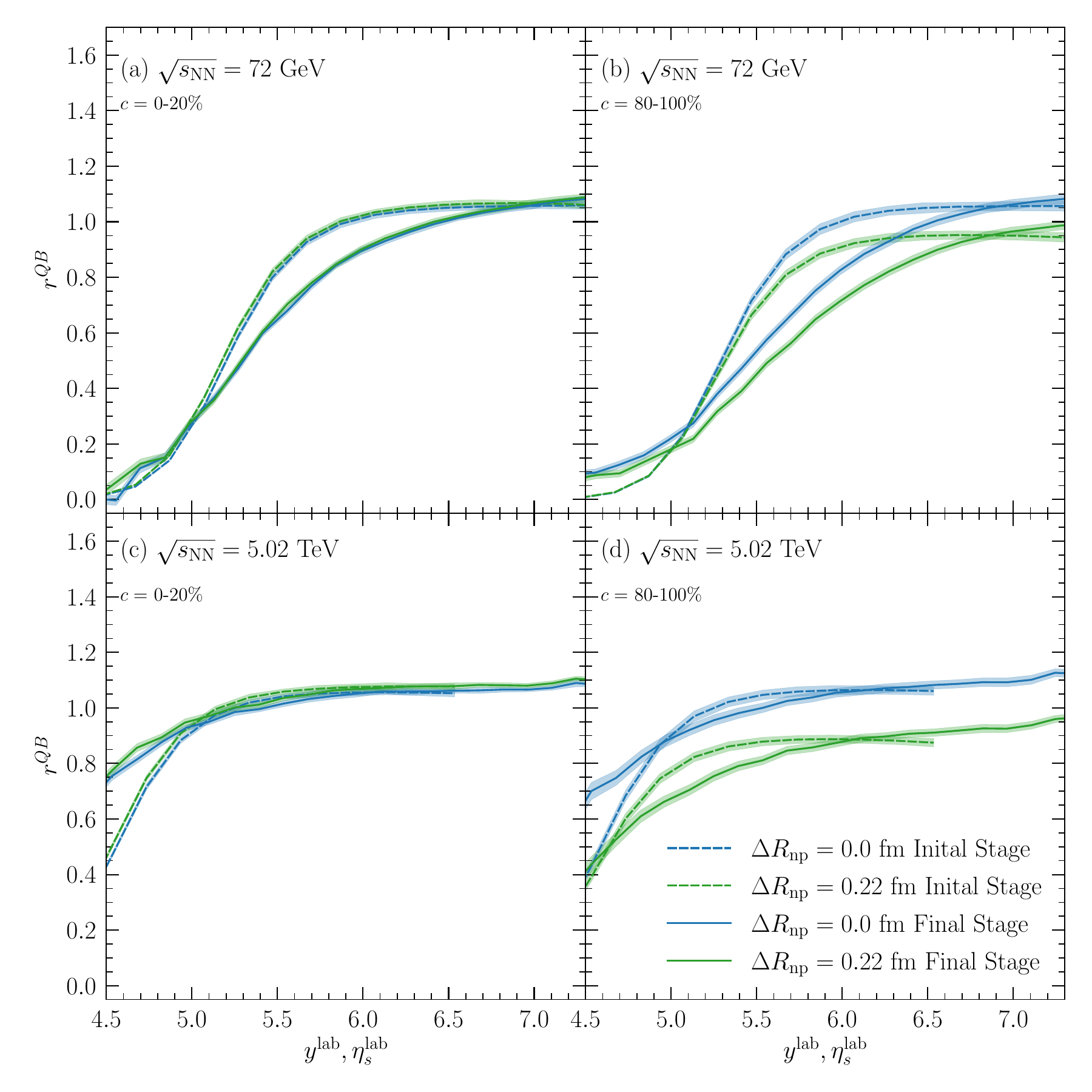}
    \caption{The ratio $r_c^{QB} =\langle dN_{Q}/dy\rangle_{\rm ev}(y, c)/\langle dN_{B}/dy\rangle_{\rm ev}(y, c) \times Z/A$ for fixed target (panels (a) and (b) and collider mode (panels (c) and (d)) as a function of the rapidity in the laboratory frame at initial stage ($\eta^{\rm lab}_s$) and final stage ($y^{\rm lab}$). The left panels represents the central collisions and the right panel are the peripheral collisions. The dashed lines corresponds to the initial stage and the solid lines to the final stage.
    }
    \label{fig:SMsmallr}
\end{figure} 
Even though the deposited net baryon and net electric charge undergo a dynamical evolution during the hydrodynamic and hadron transport stages of the collision, most dynamical effects are canceled in their ratio. 

In Fig.~\ref{fig:SMsmallr}, we show the initial and final stage $r_c^{QB} =\langle dN_{Q}/dy\rangle_{\rm ev}(y, c)/\langle dN_{B}/dy\rangle_{\rm ev}(y, c) \times Z/A$ ratios, where $Z = 82$ and $A=208$, as a function of the rapidity for central  (panel (a) and (c)) and peripheral (panel (b) and (d)) collisions for $\Delta R_{\rm np} = 0$~fm and $\Delta R_{\rm np} = 0.22$~fm. We observe that these ratios conserve the effect of the neutron skin from the initial stage (dashed lines) to the final stage (solid lines). The rapidity distributions differ between the two stages but the overall amplitude of the neutron skin effect is conserved, both for central and peripheral collisions. The double ratio Eq.~\eqref{eq:Rdef} is constructed by integrating over the relevant rapidity window and taking the ratio of peripheral to central $r^{QB}$, isolating the neutron skin effect.

\bibliography{biblio, non-inspires}

\begin{thebibliography}{56}%
\makeatletter
\providecommand \@ifxundefined [1]{%
 \@ifx{#1\undefined}
}%
\providecommand \@ifnum [1]{%
 \ifnum #1\expandafter \@firstoftwo
 \else \expandafter \@secondoftwo
 \fi
}%
\providecommand \@ifx [1]{%
 \ifx #1\expandafter \@firstoftwo
 \else \expandafter \@secondoftwo
 \fi
}%
\providecommand \natexlab [1]{#1}%
\providecommand \enquote  [1]{``#1''}%
\providecommand \bibnamefont  [1]{#1}%
\providecommand \bibfnamefont [1]{#1}%
\providecommand \citenamefont [1]{#1}%
\providecommand \href@noop [0]{\@secondoftwo}%
\providecommand \href [0]{\begingroup \@sanitize@url \@href}%
\providecommand \@href[1]{\@@startlink{#1}\@@href}%
\providecommand \@@href[1]{\endgroup#1\@@endlink}%
\providecommand \@sanitize@url [0]{\catcode `\\12\catcode `\$12\catcode `\&12\catcode `\#12\catcode `\^12\catcode `\_12\catcode `\%12\relax}%
\providecommand \@@startlink[1]{}%
\providecommand \@@endlink[0]{}%
\providecommand \url  [0]{\begingroup\@sanitize@url \@url }%
\providecommand \@url [1]{\endgroup\@href {#1}{\urlprefix }}%
\providecommand \urlprefix  [0]{URL }%
\providecommand \Eprint [0]{\href }%
\providecommand \doibase [0]{https://doi.org/}%
\providecommand \selectlanguage [0]{\@gobble}%
\providecommand \bibinfo  [0]{\@secondoftwo}%
\providecommand \bibfield  [0]{\@secondoftwo}%
\providecommand \translation [1]{[#1]}%
\providecommand \BibitemOpen [0]{}%
\providecommand \bibitemStop [0]{}%
\providecommand \bibitemNoStop [0]{.\EOS\space}%
\providecommand \EOS [0]{\spacefactor3000\relax}%
\providecommand \BibitemShut  [1]{\csname bibitem#1\endcsname}%
\let\auto@bib@innerbib\@empty
\bibitem [{\citenamefont {Chen}\ \emph {et~al.}(2010)\citenamefont {Chen}, \citenamefont {Ko}, \citenamefont {Li},\ and\ \citenamefont {Xu}}]{Chen:2010qx}%
  \BibitemOpen
  \bibfield  {author} {\bibinfo {author} {\bibfnamefont {L.-W.}\ \bibnamefont {Chen}}, \bibinfo {author} {\bibfnamefont {C.~M.}\ \bibnamefont {Ko}}, \bibinfo {author} {\bibfnamefont {B.-A.}\ \bibnamefont {Li}},\ and\ \bibinfo {author} {\bibfnamefont {J.}~\bibnamefont {Xu}},\ }\bibfield  {title} {\bibinfo {title} {{Density slope of the nuclear symmetry energy from the neutron skin thickness of heavy nuclei}},\ }\href {https://doi.org/10.1103/PhysRevC.82.024321} {\bibfield  {journal} {\bibinfo  {journal} {Phys. Rev. C}\ }\textbf {\bibinfo {volume} {82}},\ \bibinfo {pages} {024321} (\bibinfo {year} {2010})},\ \Eprint {https://arxiv.org/abs/1004.4672} {arXiv:1004.4672 [nucl-th]} \BibitemShut {NoStop}%
\bibitem [{\citenamefont {Lovato}\ \emph {et~al.}(2022)\citenamefont {Lovato} \emph {et~al.}}]{Lovato:2022vgq}%
  \BibitemOpen
  \bibfield  {author} {\bibinfo {author} {\bibfnamefont {A.}~\bibnamefont {Lovato}} \emph {et~al.},\ }\bibfield  {title} {\bibinfo {title} {{Long Range Plan: Dense matter theory for heavy-ion collisions and neutron stars}},\ }\href@noop {} {\  (\bibinfo {year} {2022})},\ \Eprint {https://arxiv.org/abs/2211.02224} {arXiv:2211.02224 [nucl-th]} \BibitemShut {NoStop}%
\bibitem [{\citenamefont {Sorensen}\ \emph {et~al.}(2024)\citenamefont {Sorensen} \emph {et~al.}}]{Sorensen:2023zkk}%
  \BibitemOpen
  \bibfield  {author} {\bibinfo {author} {\bibfnamefont {A.}~\bibnamefont {Sorensen}} \emph {et~al.},\ }\bibfield  {title} {\bibinfo {title} {{Dense nuclear matter equation of state from heavy-ion collisions}},\ }\href {https://doi.org/10.1016/j.ppnp.2023.104080} {\bibfield  {journal} {\bibinfo  {journal} {Prog. Part. Nucl. Phys.}\ }\textbf {\bibinfo {volume} {134}},\ \bibinfo {pages} {104080} (\bibinfo {year} {2024})},\ \Eprint {https://arxiv.org/abs/2301.13253} {arXiv:2301.13253 [nucl-th]} \BibitemShut {NoStop}%
\bibitem [{\citenamefont {Drischler}\ \emph {et~al.}(2020)\citenamefont {Drischler}, \citenamefont {Furnstahl}, \citenamefont {Melendez},\ and\ \citenamefont {Phillips}}]{Drischler:2020hwi}%
  \BibitemOpen
  \bibfield  {author} {\bibinfo {author} {\bibfnamefont {C.}~\bibnamefont {Drischler}}, \bibinfo {author} {\bibfnamefont {R.~J.}\ \bibnamefont {Furnstahl}}, \bibinfo {author} {\bibfnamefont {J.~A.}\ \bibnamefont {Melendez}},\ and\ \bibinfo {author} {\bibfnamefont {D.~R.}\ \bibnamefont {Phillips}},\ }\bibfield  {title} {\bibinfo {title} {{How Well Do We Know the Neutron-Matter Equation of State at the Densities Inside Neutron Stars? A Bayesian Approach with Correlated Uncertainties}},\ }\href {https://doi.org/10.1103/PhysRevLett.125.202702} {\bibfield  {journal} {\bibinfo  {journal} {Phys. Rev. Lett.}\ }\textbf {\bibinfo {volume} {125}},\ \bibinfo {pages} {202702} (\bibinfo {year} {2020})},\ \Eprint {https://arxiv.org/abs/2004.07232} {arXiv:2004.07232 [nucl-th]} \BibitemShut {NoStop}%
\bibitem [{\citenamefont {Fricke}\ \emph {et~al.}(1995)\citenamefont {Fricke}, \citenamefont {Bernhardt}, \citenamefont {Heilig}, \citenamefont {Schaller}, \citenamefont {Schellenberg}, \citenamefont {Shera},\ and\ \citenamefont {de~Jager}}]{Fricke:1995zz}%
  \BibitemOpen
  \bibfield  {author} {\bibinfo {author} {\bibfnamefont {G.}~\bibnamefont {Fricke}}, \bibinfo {author} {\bibfnamefont {C.}~\bibnamefont {Bernhardt}}, \bibinfo {author} {\bibfnamefont {K.}~\bibnamefont {Heilig}}, \bibinfo {author} {\bibfnamefont {L.~A.}\ \bibnamefont {Schaller}}, \bibinfo {author} {\bibfnamefont {L.}~\bibnamefont {Schellenberg}}, \bibinfo {author} {\bibfnamefont {E.~B.}\ \bibnamefont {Shera}},\ and\ \bibinfo {author} {\bibfnamefont {C.~W.}\ \bibnamefont {de~Jager}},\ }\bibfield  {title} {\bibinfo {title} {{Nuclear Ground State Charge Radii from Electromagnetic Interactions}},\ }\href {https://doi.org/10.1006/adnd.1995.1007} {\bibfield  {journal} {\bibinfo  {journal} {Atom. Data Nucl. Data Tabl.}\ }\textbf {\bibinfo {volume} {60}},\ \bibinfo {pages} {177} (\bibinfo {year} {1995})}\BibitemShut {NoStop}%
\bibitem [{\citenamefont {Thiel}\ \emph {et~al.}(2019)\citenamefont {Thiel}, \citenamefont {Sfienti}, \citenamefont {Piekarewicz}, \citenamefont {Horowitz},\ and\ \citenamefont {Vanderhaeghen}}]{Thiel:2019tkm}%
  \BibitemOpen
  \bibfield  {author} {\bibinfo {author} {\bibfnamefont {M.}~\bibnamefont {Thiel}}, \bibinfo {author} {\bibfnamefont {C.}~\bibnamefont {Sfienti}}, \bibinfo {author} {\bibfnamefont {J.}~\bibnamefont {Piekarewicz}}, \bibinfo {author} {\bibfnamefont {C.~J.}\ \bibnamefont {Horowitz}},\ and\ \bibinfo {author} {\bibfnamefont {M.}~\bibnamefont {Vanderhaeghen}},\ }\bibfield  {title} {\bibinfo {title} {{Neutron skins of atomic nuclei: per aspera ad astra}},\ }\href {https://doi.org/10.1088/1361-6471/ab2c6d} {\bibfield  {journal} {\bibinfo  {journal} {J. Phys. G}\ }\textbf {\bibinfo {volume} {46}},\ \bibinfo {pages} {093003} (\bibinfo {year} {2019})},\ \Eprint {https://arxiv.org/abs/1904.12269} {arXiv:1904.12269 [nucl-ex]} \BibitemShut {NoStop}%
\bibitem [{\citenamefont {Adhikari}\ \emph {et~al.}(2021)\citenamefont {Adhikari} \emph {et~al.}}]{PREX:2021umo}%
  \BibitemOpen
  \bibfield  {author} {\bibinfo {author} {\bibfnamefont {D.}~\bibnamefont {Adhikari}} \emph {et~al.} (\bibinfo {collaboration} {PREX}),\ }\bibfield  {title} {\bibinfo {title} {{Accurate Determination of the Neutron Skin Thickness of $^{208}$Pb through Parity-Violation in Electron Scattering}},\ }\href {https://doi.org/10.1103/PhysRevLett.126.172502} {\bibfield  {journal} {\bibinfo  {journal} {Phys. Rev. Lett.}\ }\textbf {\bibinfo {volume} {126}},\ \bibinfo {pages} {172502} (\bibinfo {year} {2021})},\ \Eprint {https://arxiv.org/abs/2102.10767} {arXiv:2102.10767 [nucl-ex]} \BibitemShut {NoStop}%
\bibitem [{\citenamefont {Adhikari}\ \emph {et~al.}(2022)\citenamefont {Adhikari} \emph {et~al.}}]{CREX:2022kgg}%
  \BibitemOpen
  \bibfield  {author} {\bibinfo {author} {\bibfnamefont {D.}~\bibnamefont {Adhikari}} \emph {et~al.} (\bibinfo {collaboration} {CREX}),\ }\bibfield  {title} {\bibinfo {title} {{Precision Determination of the Neutral Weak Form Factor of Ca48}},\ }\href {https://doi.org/10.1103/PhysRevLett.129.042501} {\bibfield  {journal} {\bibinfo  {journal} {Phys. Rev. Lett.}\ }\textbf {\bibinfo {volume} {129}},\ \bibinfo {pages} {042501} (\bibinfo {year} {2022})},\ \Eprint {https://arxiv.org/abs/2205.11593} {arXiv:2205.11593 [nucl-ex]} \BibitemShut {NoStop}%
\bibitem [{\citenamefont {Zenihiro}\ \emph {et~al.}(2010)\citenamefont {Zenihiro} \emph {et~al.}}]{Zenihiro:2010zz}%
  \BibitemOpen
  \bibfield  {author} {\bibinfo {author} {\bibfnamefont {J.}~\bibnamefont {Zenihiro}} \emph {et~al.},\ }\bibfield  {title} {\bibinfo {title} {{Neutron density distributions of Pb-204, Pb-206, Pb-208 deduced via proton elastic scattering at Ep=295 MeV}},\ }\href {https://doi.org/10.1103/PhysRevC.82.044611} {\bibfield  {journal} {\bibinfo  {journal} {Phys. Rev. C}\ }\textbf {\bibinfo {volume} {82}},\ \bibinfo {pages} {044611} (\bibinfo {year} {2010})}\BibitemShut {NoStop}%
\bibitem [{\citenamefont {Gils}\ \emph {et~al.}(1980)\citenamefont {Gils}, \citenamefont {Friedman}, \citenamefont {Rebel}, \citenamefont {Buschmann}, \citenamefont {Zagromski}, \citenamefont {Klewe-Nebenius}, \citenamefont {Neumann}, \citenamefont {Pesl},\ and\ \citenamefont {Bechtold}}]{Gils:1980zza}%
  \BibitemOpen
  \bibfield  {author} {\bibinfo {author} {\bibfnamefont {H.~J.}\ \bibnamefont {Gils}}, \bibinfo {author} {\bibfnamefont {E.}~\bibnamefont {Friedman}}, \bibinfo {author} {\bibfnamefont {H.}~\bibnamefont {Rebel}}, \bibinfo {author} {\bibfnamefont {J.}~\bibnamefont {Buschmann}}, \bibinfo {author} {\bibfnamefont {S.}~\bibnamefont {Zagromski}}, \bibinfo {author} {\bibfnamefont {H.}~\bibnamefont {Klewe-Nebenius}}, \bibinfo {author} {\bibfnamefont {B.}~\bibnamefont {Neumann}}, \bibinfo {author} {\bibfnamefont {R.}~\bibnamefont {Pesl}},\ and\ \bibinfo {author} {\bibfnamefont {G.}~\bibnamefont {Bechtold}},\ }\bibfield  {title} {\bibinfo {title} {{Nuclear sizes of Ca-40, Ca-42, Ca-44, Ca-48 from elastic scattering of 104 MeV alpha particles. 1. Experimental results and optical potentials}},\ }\href {https://doi.org/10.1103/PhysRevC.21.1239} {\bibfield  {journal} {\bibinfo  {journal} {Phys. Rev. C}\ }\textbf {\bibinfo {volume} {21}},\ \bibinfo {pages} {1239} (\bibinfo {year} {1980})}\BibitemShut {NoStop}%
\bibitem [{\citenamefont {Friedman}(2012)}]{Friedman:2012pa}%
  \BibitemOpen
  \bibfield  {author} {\bibinfo {author} {\bibfnamefont {E.}~\bibnamefont {Friedman}},\ }\bibfield  {title} {\bibinfo {title} {{Neutron skins of $^{208}$Pb and $^{48}$Ca from pionic probes}},\ }\href {https://doi.org/10.1016/j.nuclphysa.2012.09.007} {\bibfield  {journal} {\bibinfo  {journal} {Nucl. Phys. A}\ }\textbf {\bibinfo {volume} {896}},\ \bibinfo {pages} {46} (\bibinfo {year} {2012})},\ \Eprint {https://arxiv.org/abs/1209.6168} {arXiv:1209.6168 [nucl-ex]} \BibitemShut {NoStop}%
\bibitem [{\citenamefont {Trzcinska}\ \emph {et~al.}(2001)\citenamefont {Trzcinska}, \citenamefont {Jastrzebski}, \citenamefont {Lubinski}, \citenamefont {Hartmann}, \citenamefont {Schmidt}, \citenamefont {von Egidy},\ and\ \citenamefont {Klos}}]{Trzcinska:2001sy}%
  \BibitemOpen
  \bibfield  {author} {\bibinfo {author} {\bibfnamefont {A.}~\bibnamefont {Trzcinska}}, \bibinfo {author} {\bibfnamefont {J.}~\bibnamefont {Jastrzebski}}, \bibinfo {author} {\bibfnamefont {P.}~\bibnamefont {Lubinski}}, \bibinfo {author} {\bibfnamefont {F.~J.}\ \bibnamefont {Hartmann}}, \bibinfo {author} {\bibfnamefont {R.}~\bibnamefont {Schmidt}}, \bibinfo {author} {\bibfnamefont {T.}~\bibnamefont {von Egidy}},\ and\ \bibinfo {author} {\bibfnamefont {B.}~\bibnamefont {Klos}},\ }\bibfield  {title} {\bibinfo {title} {{Neutron density distributions deduced from anti-protonic atoms}},\ }\href {https://doi.org/10.1103/PhysRevLett.87.082501} {\bibfield  {journal} {\bibinfo  {journal} {Phys. Rev. Lett.}\ }\textbf {\bibinfo {volume} {87}},\ \bibinfo {pages} {082501} (\bibinfo {year} {2001})}\BibitemShut {NoStop}%
\bibitem [{\citenamefont {Tamii}\ \emph {et~al.}(2011)\citenamefont {Tamii} \emph {et~al.}}]{Tamii:2011pv}%
  \BibitemOpen
  \bibfield  {author} {\bibinfo {author} {\bibfnamefont {A.}~\bibnamefont {Tamii}} \emph {et~al.},\ }\bibfield  {title} {\bibinfo {title} {{Complete electric dipole response and the neutron skin in 208Pb}},\ }\href {https://doi.org/10.1103/PhysRevLett.107.062502} {\bibfield  {journal} {\bibinfo  {journal} {Phys. Rev. Lett.}\ }\textbf {\bibinfo {volume} {107}},\ \bibinfo {pages} {062502} (\bibinfo {year} {2011})},\ \Eprint {https://arxiv.org/abs/1104.5431} {arXiv:1104.5431 [nucl-ex]} \BibitemShut {NoStop}%
\bibitem [{\citenamefont {Poltoratska}\ \emph {et~al.}(2012)\citenamefont {Poltoratska} \emph {et~al.}}]{Poltoratska:2012nf}%
  \BibitemOpen
  \bibfield  {author} {\bibinfo {author} {\bibfnamefont {I.}~\bibnamefont {Poltoratska}} \emph {et~al.},\ }\bibfield  {title} {\bibinfo {title} {{Pygmy dipole resonance in $^{208}Pb$}},\ }\href {https://doi.org/10.1103/PhysRevC.85.041304} {\bibfield  {journal} {\bibinfo  {journal} {Phys. Rev. C}\ }\textbf {\bibinfo {volume} {85}},\ \bibinfo {pages} {041304} (\bibinfo {year} {2012})},\ \Eprint {https://arxiv.org/abs/1203.2155} {arXiv:1203.2155 [nucl-ex]} \BibitemShut {NoStop}%
\bibitem [{\citenamefont {Tarbert}\ \emph {et~al.}(2014)\citenamefont {Tarbert} \emph {et~al.}}]{Tarbert:2013jze}%
  \BibitemOpen
  \bibfield  {author} {\bibinfo {author} {\bibfnamefont {C.~M.}\ \bibnamefont {Tarbert}} \emph {et~al.},\ }\bibfield  {title} {\bibinfo {title} {{Neutron skin of $^{208}$Pb from Coherent Pion Photoproduction}},\ }\href {https://doi.org/10.1103/PhysRevLett.112.242502} {\bibfield  {journal} {\bibinfo  {journal} {Phys. Rev. Lett.}\ }\textbf {\bibinfo {volume} {112}},\ \bibinfo {pages} {242502} (\bibinfo {year} {2014})},\ \Eprint {https://arxiv.org/abs/1311.0168} {arXiv:1311.0168 [nucl-ex]} \BibitemShut {NoStop}%
\bibitem [{\citenamefont {Hu}\ \emph {et~al.}(2022)\citenamefont {Hu} \emph {et~al.}}]{Hu:2021trw}%
  \BibitemOpen
  \bibfield  {author} {\bibinfo {author} {\bibfnamefont {B.}~\bibnamefont {Hu}} \emph {et~al.},\ }\bibfield  {title} {\bibinfo {title} {{Ab initio predictions link the neutron skin of $^{208}$Pb to nuclear forces}},\ }\href {https://doi.org/10.1038/s41567-023-02324-9} {\bibfield  {journal} {\bibinfo  {journal} {Nature Phys.}\ }\textbf {\bibinfo {volume} {18}},\ \bibinfo {pages} {1196} (\bibinfo {year} {2022})},\ \Eprint {https://arxiv.org/abs/2112.01125} {arXiv:2112.01125 [nucl-th]} \BibitemShut {NoStop}%
\bibitem [{\citenamefont {Lattimer}\ \emph {et~al.}(1991)\citenamefont {Lattimer}, \citenamefont {Prakash}, \citenamefont {Pethick},\ and\ \citenamefont {Haensel}}]{Lattimer:1991ib}%
  \BibitemOpen
  \bibfield  {author} {\bibinfo {author} {\bibfnamefont {J.~M.}\ \bibnamefont {Lattimer}}, \bibinfo {author} {\bibfnamefont {M.}~\bibnamefont {Prakash}}, \bibinfo {author} {\bibfnamefont {C.~J.}\ \bibnamefont {Pethick}},\ and\ \bibinfo {author} {\bibfnamefont {P.}~\bibnamefont {Haensel}},\ }\bibfield  {title} {\bibinfo {title} {{Direct URCA process in neutron stars}},\ }\href {https://doi.org/10.1103/PhysRevLett.66.2701} {\bibfield  {journal} {\bibinfo  {journal} {Phys. Rev. Lett.}\ }\textbf {\bibinfo {volume} {66}},\ \bibinfo {pages} {2701} (\bibinfo {year} {1991})}\BibitemShut {NoStop}%
\bibitem [{\citenamefont {Thapa}\ and\ \citenamefont {Sinha}(2022)}]{Thapa:2022zkr}%
  \BibitemOpen
  \bibfield  {author} {\bibinfo {author} {\bibfnamefont {V.~B.}\ \bibnamefont {Thapa}}\ and\ \bibinfo {author} {\bibfnamefont {M.}~\bibnamefont {Sinha}},\ }\bibfield  {title} {\bibinfo {title} {{Direct URCA process in light of PREX-2}},\ }\href@noop {} {\  (\bibinfo {year} {2022})},\ \Eprint {https://arxiv.org/abs/2203.02272} {arXiv:2203.02272 [nucl-th]} \BibitemShut {NoStop}%
\bibitem [{\citenamefont {Cromartie}\ \emph {et~al.}(2019)\citenamefont {Cromartie} \emph {et~al.}}]{NANOGrav:2019jur}%
  \BibitemOpen
  \bibfield  {author} {\bibinfo {author} {\bibfnamefont {H.~T.}\ \bibnamefont {Cromartie}} \emph {et~al.} (\bibinfo {collaboration} {NANOGrav}),\ }\bibfield  {title} {\bibinfo {title} {{Relativistic Shapiro delay measurements of an extremely massive millisecond pulsar}},\ }\href {https://doi.org/10.1038/s41550-019-0880-2} {\bibfield  {journal} {\bibinfo  {journal} {Nature Astron.}\ }\textbf {\bibinfo {volume} {4}},\ \bibinfo {pages} {72} (\bibinfo {year} {2019})},\ \Eprint {https://arxiv.org/abs/1904.06759} {arXiv:1904.06759 [astro-ph.HE]} \BibitemShut {NoStop}%
\bibitem [{\citenamefont {Fonseca}\ \emph {et~al.}(2021)\citenamefont {Fonseca} \emph {et~al.}}]{Fonseca:2021wxt}%
  \BibitemOpen
  \bibfield  {author} {\bibinfo {author} {\bibfnamefont {E.}~\bibnamefont {Fonseca}} \emph {et~al.},\ }\bibfield  {title} {\bibinfo {title} {{Refined Mass and Geometric Measurements of the High-mass PSR J0740+6620}},\ }\href {https://doi.org/10.3847/2041-8213/ac03b8} {\bibfield  {journal} {\bibinfo  {journal} {Astrophys. J. Lett.}\ }\textbf {\bibinfo {volume} {915}},\ \bibinfo {pages} {L12} (\bibinfo {year} {2021})},\ \Eprint {https://arxiv.org/abs/2104.00880} {arXiv:2104.00880 [astro-ph.HE]} \BibitemShut {NoStop}%
\bibitem [{\citenamefont {Miller}\ \emph {et~al.}(2021)\citenamefont {Miller} \emph {et~al.}}]{Miller:2021qha}%
  \BibitemOpen
  \bibfield  {author} {\bibinfo {author} {\bibfnamefont {M.~C.}\ \bibnamefont {Miller}} \emph {et~al.},\ }\bibfield  {title} {\bibinfo {title} {{The Radius of PSR J0740+6620 from NICER and XMM-Newton Data}},\ }\href {https://doi.org/10.3847/2041-8213/ac089b} {\bibfield  {journal} {\bibinfo  {journal} {Astrophys. J. Lett.}\ }\textbf {\bibinfo {volume} {918}},\ \bibinfo {pages} {L28} (\bibinfo {year} {2021})},\ \Eprint {https://arxiv.org/abs/2105.06979} {arXiv:2105.06979 [astro-ph.HE]} \BibitemShut {NoStop}%
\bibitem [{\citenamefont {Giacalone}\ \emph {et~al.}(2023)\citenamefont {Giacalone}, \citenamefont {Nijs},\ and\ \citenamefont {van~der Schee}}]{Giacalone:2023cet}%
  \BibitemOpen
  \bibfield  {author} {\bibinfo {author} {\bibfnamefont {G.}~\bibnamefont {Giacalone}}, \bibinfo {author} {\bibfnamefont {G.}~\bibnamefont {Nijs}},\ and\ \bibinfo {author} {\bibfnamefont {W.}~\bibnamefont {van~der Schee}},\ }\bibfield  {title} {\bibinfo {title} {{Determination of the Neutron Skin of Pb208 from Ultrarelativistic Nuclear Collisions}},\ }\href {https://doi.org/10.1103/PhysRevLett.131.202302} {\bibfield  {journal} {\bibinfo  {journal} {Phys. Rev. Lett.}\ }\textbf {\bibinfo {volume} {131}},\ \bibinfo {pages} {202302} (\bibinfo {year} {2023})},\ \Eprint {https://arxiv.org/abs/2305.00015} {arXiv:2305.00015 [nucl-th]} \BibitemShut {NoStop}%
\bibitem [{\citenamefont {Liu}\ \emph {et~al.}(2023)\citenamefont {Liu}, \citenamefont {Xu},\ and\ \citenamefont {Peng}}]{Liu:2023qeq}%
  \BibitemOpen
  \bibfield  {author} {\bibinfo {author} {\bibfnamefont {L.-M.}\ \bibnamefont {Liu}}, \bibinfo {author} {\bibfnamefont {J.}~\bibnamefont {Xu}},\ and\ \bibinfo {author} {\bibfnamefont {G.-X.}\ \bibnamefont {Peng}},\ }\bibfield  {title} {\bibinfo {title} {{Measuring deformed neutron skin with free spectator nucleons in relativistic heavy-ion collisions}},\ }\href {https://doi.org/10.1016/j.physletb.2023.137701} {\bibfield  {journal} {\bibinfo  {journal} {Phys. Lett. B}\ }\textbf {\bibinfo {volume} {838}},\ \bibinfo {pages} {137701} (\bibinfo {year} {2023})},\ \Eprint {https://arxiv.org/abs/2301.07893} {arXiv:2301.07893 [nucl-th]} \BibitemShut {NoStop}%
\bibitem [{\citenamefont {Zhang}\ \emph {et~al.}(2025)\citenamefont {Zhang}, \citenamefont {Wang}, \citenamefont {Wang},\ and\ \citenamefont {Xing}}]{Zhang:2025raf}%
  \BibitemOpen
  \bibfield  {author} {\bibinfo {author} {\bibfnamefont {S.-L.}\ \bibnamefont {Zhang}}, \bibinfo {author} {\bibfnamefont {E.}~\bibnamefont {Wang}}, \bibinfo {author} {\bibfnamefont {X.-N.}\ \bibnamefont {Wang}},\ and\ \bibinfo {author} {\bibfnamefont {H.}~\bibnamefont {Xing}},\ }\bibfield  {title} {\bibinfo {title} {{Unraveling the neutron skin thickness through jet charge in deep inelastic scattering}},\ }\href@noop {} {\  (\bibinfo {year} {2025})},\ \Eprint {https://arxiv.org/abs/2506.10694} {arXiv:2506.10694 [hep-ph]} \BibitemShut {NoStop}%
\bibitem [{\citenamefont {Andersson}\ \emph {et~al.}(1983)\citenamefont {Andersson}, \citenamefont {Gustafson}, \citenamefont {Ingelman},\ and\ \citenamefont {Sjostrand}}]{Andersson:1983ia}%
  \BibitemOpen
  \bibfield  {author} {\bibinfo {author} {\bibfnamefont {B.}~\bibnamefont {Andersson}}, \bibinfo {author} {\bibfnamefont {G.}~\bibnamefont {Gustafson}}, \bibinfo {author} {\bibfnamefont {G.}~\bibnamefont {Ingelman}},\ and\ \bibinfo {author} {\bibfnamefont {T.}~\bibnamefont {Sjostrand}},\ }\bibfield  {title} {\bibinfo {title} {{Parton Fragmentation and String Dynamics}},\ }\href {https://doi.org/10.1016/0370-1573(83)90080-7} {\bibfield  {journal} {\bibinfo  {journal} {Phys. Rept.}\ }\textbf {\bibinfo {volume} {97}},\ \bibinfo {pages} {31} (\bibinfo {year} {1983})}\BibitemShut {NoStop}%
\bibitem [{\citenamefont {Vance}\ \emph {et~al.}(1998)\citenamefont {Vance}, \citenamefont {Gyulassy},\ and\ \citenamefont {Wang}}]{Vance:1997th}%
  \BibitemOpen
  \bibfield  {author} {\bibinfo {author} {\bibfnamefont {S.~E.}\ \bibnamefont {Vance}}, \bibinfo {author} {\bibfnamefont {M.}~\bibnamefont {Gyulassy}},\ and\ \bibinfo {author} {\bibfnamefont {X.~N.}\ \bibnamefont {Wang}},\ }\bibfield  {title} {\bibinfo {title} {{Baryon junction stopping at the SPS and RHIC via HIJING/B}},\ }\href {https://doi.org/10.1016/S0375-9474(98)00394-7} {\bibfield  {journal} {\bibinfo  {journal} {Nucl. Phys. A}\ }\textbf {\bibinfo {volume} {638}},\ \bibinfo {pages} {395C} (\bibinfo {year} {1998})},\ \Eprint {https://arxiv.org/abs/nucl-th/9802036} {arXiv:nucl-th/9802036} \BibitemShut {NoStop}%
\bibitem [{\citenamefont {Sjostrand}\ \emph {et~al.}(2006)\citenamefont {Sjostrand}, \citenamefont {Mrenna},\ and\ \citenamefont {Skands}}]{Sjostrand:2006za}%
  \BibitemOpen
  \bibfield  {author} {\bibinfo {author} {\bibfnamefont {T.}~\bibnamefont {Sjostrand}}, \bibinfo {author} {\bibfnamefont {S.}~\bibnamefont {Mrenna}},\ and\ \bibinfo {author} {\bibfnamefont {P.~Z.}\ \bibnamefont {Skands}},\ }\bibfield  {title} {\bibinfo {title} {{PYTHIA 6.4 Physics and Manual}},\ }\href {https://doi.org/10.1088/1126-6708/2006/05/026} {\bibfield  {journal} {\bibinfo  {journal} {JHEP}\ }\textbf {\bibinfo {volume} {05}},\ \bibinfo {pages} {026}},\ \Eprint {https://arxiv.org/abs/hep-ph/0603175} {arXiv:hep-ph/0603175} \BibitemShut {NoStop}%
\bibitem [{\citenamefont {Bahr}\ \emph {et~al.}(2008)\citenamefont {Bahr} \emph {et~al.}}]{Bahr:2008pv}%
  \BibitemOpen
  \bibfield  {author} {\bibinfo {author} {\bibfnamefont {M.}~\bibnamefont {Bahr}} \emph {et~al.},\ }\bibfield  {title} {\bibinfo {title} {{Herwig++ Physics and Manual}},\ }\href {https://doi.org/10.1140/epjc/s10052-008-0798-9} {\bibfield  {journal} {\bibinfo  {journal} {Eur. Phys. J. C}\ }\textbf {\bibinfo {volume} {58}},\ \bibinfo {pages} {639} (\bibinfo {year} {2008})},\ \Eprint {https://arxiv.org/abs/0803.0883} {arXiv:0803.0883 [hep-ph]} \BibitemShut {NoStop}%
\bibitem [{\citenamefont {Pratt}(2024)}]{Pratt:2023pee}%
  \BibitemOpen
  \bibfield  {author} {\bibinfo {author} {\bibfnamefont {S.}~\bibnamefont {Pratt}},\ }\bibfield  {title} {\bibinfo {title} {{Baryon transport in color flux tubes}},\ }\href {https://doi.org/10.1103/PhysRevC.109.044910} {\bibfield  {journal} {\bibinfo  {journal} {Phys. Rev. C}\ }\textbf {\bibinfo {volume} {109}},\ \bibinfo {pages} {044910} (\bibinfo {year} {2024})},\ \Eprint {https://arxiv.org/abs/2311.17906} {arXiv:2311.17906 [hep-ph]} \BibitemShut {NoStop}%
\bibitem [{\citenamefont {Pihan}\ \emph {et~al.}(2024)\citenamefont {Pihan}, \citenamefont {Monnai}, \citenamefont {Schenke},\ and\ \citenamefont {Shen}}]{Pihan:2024lxw}%
  \BibitemOpen
  \bibfield  {author} {\bibinfo {author} {\bibfnamefont {G.}~\bibnamefont {Pihan}}, \bibinfo {author} {\bibfnamefont {A.}~\bibnamefont {Monnai}}, \bibinfo {author} {\bibfnamefont {B.}~\bibnamefont {Schenke}},\ and\ \bibinfo {author} {\bibfnamefont {C.}~\bibnamefont {Shen}},\ }\bibfield  {title} {\bibinfo {title} {{Unveiling Baryon Charge Carriers through Charge Stopping in Isobar Collisions}},\ }\href {https://doi.org/10.1103/PhysRevLett.133.182301} {\bibfield  {journal} {\bibinfo  {journal} {Phys. Rev. Lett.}\ }\textbf {\bibinfo {volume} {133}},\ \bibinfo {pages} {182301} (\bibinfo {year} {2024})},\ \Eprint {https://arxiv.org/abs/2405.19439} {arXiv:2405.19439 [nucl-th]} \BibitemShut {NoStop}%
\bibitem [{\citenamefont {Pihan}\ \emph {et~al.}(2025)\citenamefont {Pihan}, \citenamefont {Monnai}, \citenamefont {Schenke},\ and\ \citenamefont {Shen}}]{pihan_2025_17193964}%
  \BibitemOpen
  \bibfield  {author} {\bibinfo {author} {\bibfnamefont {G.}~\bibnamefont {Pihan}}, \bibinfo {author} {\bibfnamefont {A.}~\bibnamefont {Monnai}}, \bibinfo {author} {\bibfnamefont {B.}~\bibnamefont {Schenke}},\ and\ \bibinfo {author} {\bibfnamefont {C.}~\bibnamefont {Shen}},\ }\href {https://doi.org/10.5281/zenodo.17193964} {\bibinfo {title} {Repository for "neutron skin from conserved charge measurements at collider experiments"}} (\bibinfo {year} {2025})\BibitemShut {NoStop}%
\bibitem [{\citenamefont {Shen}\ and\ \citenamefont {Schenke}(2018)}]{Shen:2017bsr}%
  \BibitemOpen
  \bibfield  {author} {\bibinfo {author} {\bibfnamefont {C.}~\bibnamefont {Shen}}\ and\ \bibinfo {author} {\bibfnamefont {B.}~\bibnamefont {Schenke}},\ }\bibfield  {title} {\bibinfo {title} {{Dynamical initial state model for relativistic heavy-ion collisions}},\ }\href {https://doi.org/10.1103/PhysRevC.97.024907} {\bibfield  {journal} {\bibinfo  {journal} {Phys. Rev. C}\ }\textbf {\bibinfo {volume} {97}},\ \bibinfo {pages} {024907} (\bibinfo {year} {2018})},\ \Eprint {https://arxiv.org/abs/1710.00881} {arXiv:1710.00881 [nucl-th]} \BibitemShut {NoStop}%
\bibitem [{\citenamefont {Shen}\ and\ \citenamefont {Schenke}(2022)}]{Shen:2022oyg}%
  \BibitemOpen
  \bibfield  {author} {\bibinfo {author} {\bibfnamefont {C.}~\bibnamefont {Shen}}\ and\ \bibinfo {author} {\bibfnamefont {B.}~\bibnamefont {Schenke}},\ }\bibfield  {title} {\bibinfo {title} {{Longitudinal dynamics and particle production in relativistic nuclear collisions}},\ }\href {https://doi.org/10.1103/PhysRevC.105.064905} {\bibfield  {journal} {\bibinfo  {journal} {Phys. Rev. C}\ }\textbf {\bibinfo {volume} {105}},\ \bibinfo {pages} {064905} (\bibinfo {year} {2022})},\ \Eprint {https://arxiv.org/abs/2203.04685} {arXiv:2203.04685 [nucl-th]} \BibitemShut {NoStop}%
\bibitem [{\citenamefont {Montanet}\ \emph {et~al.}(1980)\citenamefont {Montanet}, \citenamefont {Rossi},\ and\ \citenamefont {Veneziano}}]{Montanet:1980te}%
  \BibitemOpen
  \bibfield  {author} {\bibinfo {author} {\bibfnamefont {L.}~\bibnamefont {Montanet}}, \bibinfo {author} {\bibfnamefont {G.~C.}\ \bibnamefont {Rossi}},\ and\ \bibinfo {author} {\bibfnamefont {G.}~\bibnamefont {Veneziano}},\ }\bibfield  {title} {\bibinfo {title} {{Baryonium Physics}},\ }\href {https://doi.org/10.1016/0370-1573(80)90161-1} {\bibfield  {journal} {\bibinfo  {journal} {Phys. Rept.}\ }\textbf {\bibinfo {volume} {63}},\ \bibinfo {pages} {149} (\bibinfo {year} {1980})}\BibitemShut {NoStop}%
\bibitem [{\citenamefont {Kharzeev}(1996)}]{Kharzeev:1996sq}%
  \BibitemOpen
  \bibfield  {author} {\bibinfo {author} {\bibfnamefont {D.}~\bibnamefont {Kharzeev}},\ }\bibfield  {title} {\bibinfo {title} {{Can gluons trace baryon number?}},\ }\href {https://doi.org/10.1016/0370-2693(96)00435-2} {\bibfield  {journal} {\bibinfo  {journal} {Phys. Lett. B}\ }\textbf {\bibinfo {volume} {378}},\ \bibinfo {pages} {238} (\bibinfo {year} {1996})},\ \Eprint {https://arxiv.org/abs/nucl-th/9602027} {arXiv:nucl-th/9602027} \BibitemShut {NoStop}%
\bibitem [{\citenamefont {Schenke}\ \emph {et~al.}(2010)\citenamefont {Schenke}, \citenamefont {Jeon},\ and\ \citenamefont {Gale}}]{Schenke:2010nt}%
  \BibitemOpen
  \bibfield  {author} {\bibinfo {author} {\bibfnamefont {B.}~\bibnamefont {Schenke}}, \bibinfo {author} {\bibfnamefont {S.}~\bibnamefont {Jeon}},\ and\ \bibinfo {author} {\bibfnamefont {C.}~\bibnamefont {Gale}},\ }\bibfield  {title} {\bibinfo {title} {{(3+1)D hydrodynamic simulation of relativistic heavy-ion collisions}},\ }\href {https://doi.org/10.1103/PhysRevC.82.014903} {\bibfield  {journal} {\bibinfo  {journal} {Phys. Rev. C}\ }\textbf {\bibinfo {volume} {82}},\ \bibinfo {pages} {014903} (\bibinfo {year} {2010})},\ \Eprint {https://arxiv.org/abs/1004.1408} {arXiv:1004.1408 [hep-ph]} \BibitemShut {NoStop}%
\bibitem [{\citenamefont {Schenke}\ \emph {et~al.}(2011)\citenamefont {Schenke}, \citenamefont {Jeon},\ and\ \citenamefont {Gale}}]{Schenke:2010rr}%
  \BibitemOpen
  \bibfield  {author} {\bibinfo {author} {\bibfnamefont {B.}~\bibnamefont {Schenke}}, \bibinfo {author} {\bibfnamefont {S.}~\bibnamefont {Jeon}},\ and\ \bibinfo {author} {\bibfnamefont {C.}~\bibnamefont {Gale}},\ }\bibfield  {title} {\bibinfo {title} {{Elliptic and triangular flow in event-by-event (3+1)D viscous hydrodynamics}},\ }\href {https://doi.org/10.1103/PhysRevLett.106.042301} {\bibfield  {journal} {\bibinfo  {journal} {Phys. Rev. Lett.}\ }\textbf {\bibinfo {volume} {106}},\ \bibinfo {pages} {042301} (\bibinfo {year} {2011})},\ \Eprint {https://arxiv.org/abs/1009.3244} {arXiv:1009.3244 [hep-ph]} \BibitemShut {NoStop}%
\bibitem [{\citenamefont {Paquet}\ \emph {et~al.}(2016)\citenamefont {Paquet}, \citenamefont {Shen}, \citenamefont {Denicol}, \citenamefont {Luzum}, \citenamefont {Schenke}, \citenamefont {Jeon},\ and\ \citenamefont {Gale}}]{Paquet:2015lta}%
  \BibitemOpen
  \bibfield  {author} {\bibinfo {author} {\bibfnamefont {J.-F.}\ \bibnamefont {Paquet}}, \bibinfo {author} {\bibfnamefont {C.}~\bibnamefont {Shen}}, \bibinfo {author} {\bibfnamefont {G.~S.}\ \bibnamefont {Denicol}}, \bibinfo {author} {\bibfnamefont {M.}~\bibnamefont {Luzum}}, \bibinfo {author} {\bibfnamefont {B.}~\bibnamefont {Schenke}}, \bibinfo {author} {\bibfnamefont {S.}~\bibnamefont {Jeon}},\ and\ \bibinfo {author} {\bibfnamefont {C.}~\bibnamefont {Gale}},\ }\bibfield  {title} {\bibinfo {title} {{Production of photons in relativistic heavy-ion collisions}},\ }\href {https://doi.org/10.1103/PhysRevC.93.044906} {\bibfield  {journal} {\bibinfo  {journal} {Phys. Rev. C}\ }\textbf {\bibinfo {volume} {93}},\ \bibinfo {pages} {044906} (\bibinfo {year} {2016})},\ \Eprint {https://arxiv.org/abs/1509.06738} {arXiv:1509.06738 [hep-ph]} \BibitemShut {NoStop}%
\bibitem [{\citenamefont {Denicol}\ \emph {et~al.}(2018)\citenamefont {Denicol}, \citenamefont {Gale}, \citenamefont {Jeon}, \citenamefont {Monnai}, \citenamefont {Schenke},\ and\ \citenamefont {Shen}}]{Denicol:2018wdp}%
  \BibitemOpen
  \bibfield  {author} {\bibinfo {author} {\bibfnamefont {G.~S.}\ \bibnamefont {Denicol}}, \bibinfo {author} {\bibfnamefont {C.}~\bibnamefont {Gale}}, \bibinfo {author} {\bibfnamefont {S.}~\bibnamefont {Jeon}}, \bibinfo {author} {\bibfnamefont {A.}~\bibnamefont {Monnai}}, \bibinfo {author} {\bibfnamefont {B.}~\bibnamefont {Schenke}},\ and\ \bibinfo {author} {\bibfnamefont {C.}~\bibnamefont {Shen}},\ }\bibfield  {title} {\bibinfo {title} {{Net baryon diffusion in fluid dynamic simulations of relativistic heavy-ion collisions}},\ }\href {https://doi.org/10.1103/PhysRevC.98.034916} {\bibfield  {journal} {\bibinfo  {journal} {Phys. Rev. C}\ }\textbf {\bibinfo {volume} {98}},\ \bibinfo {pages} {034916} (\bibinfo {year} {2018})},\ \Eprint {https://arxiv.org/abs/1804.10557} {arXiv:1804.10557 [nucl-th]} \BibitemShut {NoStop}%
\bibitem [{\citenamefont {Pihan}\ \emph {et~al.}(2023)\citenamefont {Pihan}, \citenamefont {Monnai}, \citenamefont {Schenke},\ and\ \citenamefont {Shen}}]{Pihan:2023dsb}%
  \BibitemOpen
  \bibfield  {author} {\bibinfo {author} {\bibfnamefont {G.}~\bibnamefont {Pihan}}, \bibinfo {author} {\bibfnamefont {A.}~\bibnamefont {Monnai}}, \bibinfo {author} {\bibfnamefont {B.}~\bibnamefont {Schenke}},\ and\ \bibinfo {author} {\bibfnamefont {C.}~\bibnamefont {Shen}},\ }\bibfield  {title} {\bibinfo {title} {{Tracing baryon and electric charge transport in isobar collisions}}\ }(\bibinfo {year} {2023})\ \Eprint {https://arxiv.org/abs/2312.12376} {arXiv:2312.12376 [nucl-th]} \BibitemShut {NoStop}%
\bibitem [{\citenamefont {Denicol}\ \emph {et~al.}(2012)\citenamefont {Denicol}, \citenamefont {Niemi}, \citenamefont {Molnar},\ and\ \citenamefont {Rischke}}]{Denicol:2012cn}%
  \BibitemOpen
  \bibfield  {author} {\bibinfo {author} {\bibfnamefont {G.~S.}\ \bibnamefont {Denicol}}, \bibinfo {author} {\bibfnamefont {H.}~\bibnamefont {Niemi}}, \bibinfo {author} {\bibfnamefont {E.}~\bibnamefont {Molnar}},\ and\ \bibinfo {author} {\bibfnamefont {D.~H.}\ \bibnamefont {Rischke}},\ }\bibfield  {title} {\bibinfo {title} {{Derivation of transient relativistic fluid dynamics from the Boltzmann equation}},\ }\href {https://doi.org/10.1103/PhysRevD.85.114047} {\bibfield  {journal} {\bibinfo  {journal} {Phys. Rev. D}\ }\textbf {\bibinfo {volume} {85}},\ \bibinfo {pages} {114047} (\bibinfo {year} {2012})},\ \bibinfo {note} {[Erratum: Phys.Rev.D 91, 039902 (2015)]},\ \Eprint {https://arxiv.org/abs/1202.4551} {arXiv:1202.4551 [nucl-th]} \BibitemShut {NoStop}%
\bibitem [{\citenamefont {Shen}\ and\ \citenamefont {Alzhrani}(2020)}]{Shen:2020jwv}%
  \BibitemOpen
  \bibfield  {author} {\bibinfo {author} {\bibfnamefont {C.}~\bibnamefont {Shen}}\ and\ \bibinfo {author} {\bibfnamefont {S.}~\bibnamefont {Alzhrani}},\ }\bibfield  {title} {\bibinfo {title} {{Collision-geometry-based 3D initial condition for relativistic heavy-ion collisions}},\ }\href {https://doi.org/10.1103/PhysRevC.102.014909} {\bibfield  {journal} {\bibinfo  {journal} {Phys. Rev. C}\ }\textbf {\bibinfo {volume} {102}},\ \bibinfo {pages} {014909} (\bibinfo {year} {2020})},\ \Eprint {https://arxiv.org/abs/2003.05852} {arXiv:2003.05852 [nucl-th]} \BibitemShut {NoStop}%
\bibitem [{\citenamefont {Monnai}\ \emph {et~al.}(2024)\citenamefont {Monnai}, \citenamefont {Pihan}, \citenamefont {Schenke},\ and\ \citenamefont {Shen}}]{Monnai:2024pvy}%
  \BibitemOpen
  \bibfield  {author} {\bibinfo {author} {\bibfnamefont {A.}~\bibnamefont {Monnai}}, \bibinfo {author} {\bibfnamefont {G.}~\bibnamefont {Pihan}}, \bibinfo {author} {\bibfnamefont {B.}~\bibnamefont {Schenke}},\ and\ \bibinfo {author} {\bibfnamefont {C.}~\bibnamefont {Shen}},\ }\bibfield  {title} {\bibinfo {title} {{Four-dimensional QCD equation~of state with multiple chemical potentials}},\ }\href {https://doi.org/10.1103/PhysRevC.110.044905} {\bibfield  {journal} {\bibinfo  {journal} {Phys. Rev. C}\ }\textbf {\bibinfo {volume} {110}},\ \bibinfo {pages} {044905} (\bibinfo {year} {2024})},\ \Eprint {https://arxiv.org/abs/2406.11610} {arXiv:2406.11610 [nucl-th]} \BibitemShut {NoStop}%
\bibitem [{\citenamefont {Monnai}\ \emph {et~al.}(2025)\citenamefont {Monnai}, \citenamefont {Pihan}, \citenamefont {Schenke},\ and\ \citenamefont {Shen}}]{Monnai:2025nyg}%
  \BibitemOpen
  \bibfield  {author} {\bibinfo {author} {\bibfnamefont {A.}~\bibnamefont {Monnai}}, \bibinfo {author} {\bibfnamefont {G.}~\bibnamefont {Pihan}}, \bibinfo {author} {\bibfnamefont {B.}~\bibnamefont {Schenke}},\ and\ \bibinfo {author} {\bibfnamefont {C.}~\bibnamefont {Shen}},\ }\bibfield  {title} {\bibinfo {title} {{Four-dimensional QCD equation of state at finite chemical potentials}},\ }in\ \href@noop {} {\emph {\bibinfo {booktitle} {{16th Conference on Quark Confinement and the Hadron Spectrum}}}}\ (\bibinfo {year} {2025})\ \Eprint {https://arxiv.org/abs/2503.03566} {arXiv:2503.03566 [nucl-th]} \BibitemShut {NoStop}%
\bibitem [{\citenamefont {Huovinen}\ and\ \citenamefont {Petersen}(2012)}]{Huovinen:2012is}%
  \BibitemOpen
  \bibfield  {author} {\bibinfo {author} {\bibfnamefont {P.}~\bibnamefont {Huovinen}}\ and\ \bibinfo {author} {\bibfnamefont {H.}~\bibnamefont {Petersen}},\ }\bibfield  {title} {\bibinfo {title} {{Particlization in hybrid models}},\ }\href {https://doi.org/10.1140/epja/i2012-12171-9} {\bibfield  {journal} {\bibinfo  {journal} {Eur. Phys. J. A}\ }\textbf {\bibinfo {volume} {48}},\ \bibinfo {pages} {171} (\bibinfo {year} {2012})},\ \Eprint {https://arxiv.org/abs/1206.3371} {arXiv:1206.3371 [nucl-th]} \BibitemShut {NoStop}%
\bibitem [{\citenamefont {Cooper}\ and\ \citenamefont {Frye}(1974)}]{Cooper:1974mv}%
  \BibitemOpen
  \bibfield  {author} {\bibinfo {author} {\bibfnamefont {F.}~\bibnamefont {Cooper}}\ and\ \bibinfo {author} {\bibfnamefont {G.}~\bibnamefont {Frye}},\ }\bibfield  {title} {\bibinfo {title} {{Comment on the Single Particle Distribution in the Hydrodynamic and Statistical Thermodynamic Models of Multiparticle Production}},\ }\href {https://doi.org/10.1103/PhysRevD.10.186} {\bibfield  {journal} {\bibinfo  {journal} {Phys. Rev. D}\ }\textbf {\bibinfo {volume} {10}},\ \bibinfo {pages} {186} (\bibinfo {year} {1974})}\BibitemShut {NoStop}%
\bibitem [{\citenamefont {Shen}\ \emph {et~al.}(2016)\citenamefont {Shen}, \citenamefont {Qiu}, \citenamefont {Song}, \citenamefont {Bernhard}, \citenamefont {Bass},\ and\ \citenamefont {Heinz}}]{Shen:2014vra}%
  \BibitemOpen
  \bibfield  {author} {\bibinfo {author} {\bibfnamefont {C.}~\bibnamefont {Shen}}, \bibinfo {author} {\bibfnamefont {Z.}~\bibnamefont {Qiu}}, \bibinfo {author} {\bibfnamefont {H.}~\bibnamefont {Song}}, \bibinfo {author} {\bibfnamefont {J.}~\bibnamefont {Bernhard}}, \bibinfo {author} {\bibfnamefont {S.}~\bibnamefont {Bass}},\ and\ \bibinfo {author} {\bibfnamefont {U.}~\bibnamefont {Heinz}},\ }\bibfield  {title} {\bibinfo {title} {{The iEBE-VISHNU code package for relativistic heavy-ion collisions}},\ }\href {https://doi.org/10.1016/j.cpc.2015.08.039} {\bibfield  {journal} {\bibinfo  {journal} {Comput. Phys. Commun.}\ }\textbf {\bibinfo {volume} {199}},\ \bibinfo {pages} {61} (\bibinfo {year} {2016})},\ \Eprint {https://arxiv.org/abs/1409.8164} {arXiv:1409.8164 [nucl-th]} \BibitemShut {NoStop}%
\bibitem [{\citenamefont {Bass}\ \emph {et~al.}(1998)\citenamefont {Bass} \emph {et~al.}}]{Bass:1998ca}%
  \BibitemOpen
  \bibfield  {author} {\bibinfo {author} {\bibfnamefont {S.~A.}\ \bibnamefont {Bass}} \emph {et~al.},\ }\bibfield  {title} {\bibinfo {title} {{Microscopic models for ultrarelativistic heavy ion collisions}},\ }\href {https://doi.org/10.1016/S0146-6410(98)00058-1} {\bibfield  {journal} {\bibinfo  {journal} {Prog. Part. Nucl. Phys.}\ }\textbf {\bibinfo {volume} {41}},\ \bibinfo {pages} {255} (\bibinfo {year} {1998})},\ \Eprint {https://arxiv.org/abs/nucl-th/9803035} {arXiv:nucl-th/9803035} \BibitemShut {NoStop}%
\bibitem [{\citenamefont {Bleicher}\ \emph {et~al.}(1999)\citenamefont {Bleicher} \emph {et~al.}}]{Bleicher:1999xi}%
  \BibitemOpen
  \bibfield  {author} {\bibinfo {author} {\bibfnamefont {M.}~\bibnamefont {Bleicher}} \emph {et~al.},\ }\bibfield  {title} {\bibinfo {title} {{Relativistic hadron hadron collisions in the ultrarelativistic quantum molecular dynamics model}},\ }\href {https://doi.org/10.1088/0954-3899/25/9/308} {\bibfield  {journal} {\bibinfo  {journal} {J. Phys. G}\ }\textbf {\bibinfo {volume} {25}},\ \bibinfo {pages} {1859} (\bibinfo {year} {1999})},\ \Eprint {https://arxiv.org/abs/hep-ph/9909407} {arXiv:hep-ph/9909407} \BibitemShut {NoStop}%
\bibitem [{\citenamefont {Pordes}\ \emph {et~al.}(2007)\citenamefont {Pordes} \emph {et~al.}}]{Pordes:2007zzb}%
  \BibitemOpen
  \bibfield  {author} {\bibinfo {author} {\bibfnamefont {R.}~\bibnamefont {Pordes}} \emph {et~al.},\ }\bibfield  {title} {\bibinfo {title} {{The Open Science Grid}},\ }\href {https://doi.org/10.1088/1742-6596/78/1/012057} {\bibfield  {journal} {\bibinfo  {journal} {J. Phys. Conf. Ser.}\ }\textbf {\bibinfo {volume} {78}},\ \bibinfo {pages} {012057} (\bibinfo {year} {2007})}\BibitemShut {NoStop}%
\bibitem [{\citenamefont {Sfiligoi}\ \emph {et~al.}(2009)\citenamefont {Sfiligoi}, \citenamefont {Bradley}, \citenamefont {Holzman}, \citenamefont {Mhashilkar}, \citenamefont {Padhi},\ and\ \citenamefont {Wurthwrin}}]{Sfiligoi:2009cct}%
  \BibitemOpen
  \bibfield  {author} {\bibinfo {author} {\bibfnamefont {I.}~\bibnamefont {Sfiligoi}}, \bibinfo {author} {\bibfnamefont {D.~C.}\ \bibnamefont {Bradley}}, \bibinfo {author} {\bibfnamefont {B.}~\bibnamefont {Holzman}}, \bibinfo {author} {\bibfnamefont {P.}~\bibnamefont {Mhashilkar}}, \bibinfo {author} {\bibfnamefont {S.}~\bibnamefont {Padhi}},\ and\ \bibinfo {author} {\bibfnamefont {F.}~\bibnamefont {Wurthwrin}},\ }\bibfield  {title} {\bibinfo {title} {{The pilot way to Grid resources using glideinWMS}},\ }\href {https://doi.org/10.1109/CSIE.2009.950} {\bibfield  {journal} {\bibinfo  {journal} {WRI World Congress}\ }\textbf {\bibinfo {volume} {2}},\ \bibinfo {pages} {428} (\bibinfo {year} {2009})}\BibitemShut {NoStop}%
\bibitem [{\citenamefont {{OSG}}(2006)}]{OSPool}%
  \BibitemOpen
  \bibfield  {author} {\bibinfo {author} {\bibnamefont {{OSG}}},\ }\href {https://doi.org/10.21231/906P-4D78} {\bibinfo {title} {Ospool}} (\bibinfo {year} {2006})\BibitemShut {NoStop}%
\bibitem [{\citenamefont {{OSG}}(2015)}]{OSDF}%
  \BibitemOpen
  \bibfield  {author} {\bibinfo {author} {\bibnamefont {{OSG}}},\ }\href {https://doi.org/10.21231/0KVZ-VE57} {\bibinfo {title} {Open science data federation}} (\bibinfo {year} {2015})\BibitemShut {NoStop}%
\bibitem [{\citenamefont {Monnai}\ \emph {et~al.}(2026)\citenamefont {Monnai}, \citenamefont {Pihan}, \citenamefont {Schenke},\ and\ \citenamefont {Shen}}]{Monnai:2026fkp}%
  \BibitemOpen
  \bibfield  {author} {\bibinfo {author} {\bibfnamefont {A.}~\bibnamefont {Monnai}}, \bibinfo {author} {\bibfnamefont {G.}~\bibnamefont {Pihan}}, \bibinfo {author} {\bibfnamefont {B.}~\bibnamefont {Schenke}},\ and\ \bibinfo {author} {\bibfnamefont {C.}~\bibnamefont {Shen}},\ }\bibfield  {title} {\bibinfo {title} {{Signatures of QCD conductivities in heavy-ion collisions}},\ }\href@noop {} {\  (\bibinfo {year} {2026})},\ \Eprint {https://arxiv.org/abs/2601.12384} {arXiv:2601.12384 [nucl-th]} \BibitemShut {NoStop}%
\bibitem [{\citenamefont {Greif}\ \emph {et~al.}(2018)\citenamefont {Greif}, \citenamefont {Fotakis}, \citenamefont {Denicol},\ and\ \citenamefont {Greiner}}]{Greif:2017byw}%
  \BibitemOpen
  \bibfield  {author} {\bibinfo {author} {\bibfnamefont {M.}~\bibnamefont {Greif}}, \bibinfo {author} {\bibfnamefont {J.~A.}\ \bibnamefont {Fotakis}}, \bibinfo {author} {\bibfnamefont {G.~S.}\ \bibnamefont {Denicol}},\ and\ \bibinfo {author} {\bibfnamefont {C.}~\bibnamefont {Greiner}},\ }\bibfield  {title} {\bibinfo {title} {{Diffusion of conserved charges in relativistic heavy ion collisions}},\ }\href {https://doi.org/10.1103/PhysRevLett.120.242301} {\bibfield  {journal} {\bibinfo  {journal} {Phys. Rev. Lett.}\ }\textbf {\bibinfo {volume} {120}},\ \bibinfo {pages} {242301} (\bibinfo {year} {2018})},\ \Eprint {https://arxiv.org/abs/1711.08680} {arXiv:1711.08680 [hep-ph]} \BibitemShut {NoStop}%
\bibitem [{\citenamefont {Rougemont}\ \emph {et~al.}(2017)\citenamefont {Rougemont}, \citenamefont {Critelli}, \citenamefont {Noronha-Hostler}, \citenamefont {Noronha},\ and\ \citenamefont {Ratti}}]{Rougemont:2017tlu}%
  \BibitemOpen
  \bibfield  {author} {\bibinfo {author} {\bibfnamefont {R.}~\bibnamefont {Rougemont}}, \bibinfo {author} {\bibfnamefont {R.}~\bibnamefont {Critelli}}, \bibinfo {author} {\bibfnamefont {J.}~\bibnamefont {Noronha-Hostler}}, \bibinfo {author} {\bibfnamefont {J.}~\bibnamefont {Noronha}},\ and\ \bibinfo {author} {\bibfnamefont {C.}~\bibnamefont {Ratti}},\ }\bibfield  {title} {\bibinfo {title} {{Dynamical versus equilibrium properties of the QCD phase transition: A holographic perspective}},\ }\href {https://doi.org/10.1103/PhysRevD.96.014032} {\bibfield  {journal} {\bibinfo  {journal} {Phys. Rev. D}\ }\textbf {\bibinfo {volume} {96}},\ \bibinfo {pages} {014032} (\bibinfo {year} {2017})},\ \Eprint {https://arxiv.org/abs/1704.05558} {arXiv:1704.05558 [hep-ph]} \BibitemShut {NoStop}%
\end{thebibliography}%

\end{document}